\documentclass[pdflatex,sn-aps]{sn-jnl}

\usepackage{graphicx}%
\usepackage{multirow}%
\usepackage{amsmath,amssymb,amsfonts}%
\usepackage{amsthm}%
\usepackage{mathrsfs}%
\usepackage[title]{appendix}%
\usepackage{xcolor}%
\usepackage{textcomp}%
\usepackage{manyfoot}%
\usepackage{booktabs}%
\usepackage{algorithm}%
\usepackage{algorithmicx}%
\usepackage{algpseudocode}%
\usepackage{listings}%

\usepackage{colordvi}
\usepackage{color}

\begin{document}

\title[Two- and three-mode squeezing in a three-qubit entangled system]{Two- and three-mode squeezing in a three-qubit entangled system}

\author*[1]{\fnm{Joanna K.} \sur{Kalaga}}\email{j.kalaga@if.uz.zgora.pl}

\author[2]{\fnm{Jan} \sur{Pe\v{r}ina Jr.}}\email{jan.perina.jr@upol.cz}

\affil*[1]{\orgdiv{Quantum Optics and Engineering Division, Institute of Physics}, \orgname{University of Zielona G\'ora}, \orgaddress{\street{Prof.~Z.~Szafrana 4a}, \city{Zielona G\'ora}, \postcode{65-516}, \state{}, \country{Poland}}}

\affil[2]{\orgdiv{Joint Laboratory of Optics of Palack\'y University and Institute of Physics of AS CR, Faculty of Science}, \orgname{Palack\'{y} University}, \orgaddress{\street{17. listopadu 12}, \city{Olomouc}, \postcode{779~00}, \state{}, \country{Czech Republic}}}

\abstract{The states of a three-mode bosonic system with the restricted Hilbert space are discussed in the context of quantum entanglement and squeezing of quantum fluctuations. The states exhibiting non-zero tripartite entanglement are considered. Mutual relations between the two- and three-mode entanglement quantified by the corresponding negativities and the squeezing described by the corresponding principal squeeze variances are revealed. Entangled three-qubit states exhibiting squeezing are identified.}

\keywords{quantum entanglement, tripartite entanglement, quantum squeezing, three-qubit system}

\maketitle

\section{Introduction}\label{sec1}
\label{sec:intro}
Entanglement and squeezing are two fundamental concepts in quantum mechanics that are both essential to quantum information processing. Although squeezing is usually linked to continuous-variable (CV) systems with infinitely large Hilbert spaces, this concept can also be naturally applied to the cases in which the Hilbert spaces are restricted, e.g., using the method of quantum scissors.  In analogy, this concept can be defined and studied in qubit systems, where, as we demonstrate, squeezing similarly as the entanglement offers valuable insights into the quantum properties of these systems and their correlations.

Three-qubit quantum states play an important role in various areas of quantum-information processing. These states, which exhibit entanglement, are a resource for quantum computation and quantum communication \cite{HBV99,LSB04,CZZ05,K08,NC10,CH16,FML17,CG19,WCB23}. The Greenberger–Horne–Zeilinger (GHZ) states and W states play a special role. These states are characterized by the full tripartite entanglement \cite{GHZ89,DVC00,ABL01,SS07,SG08,EDZ18}. However, due to differences in the occurrence of bipartite entanglement, they have found different applications.

Entanglement of three qubits can be used in quantum cryptography to detect eavesdropping. Quantum secret sharing protocols make use of multi-qubit states. These protocols, based on the idea that a secret is shared among several parties via entanglement, ensure that no party can access the secret alone. This allows to use the entangled states in secure quantum communication protocols \cite{HBV99,CZZ05,JPL05,DZL06,WZT07,TH10,LCW24}. Three-qubit entangled states are also exploited in quantum error correction schemes \cite{YG01,TWJ08,OV10,STC17}. Both quantum computers and reliable quantum-communication protocols require protocols that protect quantum information. For instance, the authors of \cite{RDN12} demonstrated a three-qubit quantum error correction code based on super-conducting circuits.

Other compelling quantum states, besides the entangled states, are the squeezed states \cite{LPP88,Dodonov2002,Dodonov2003,Lvovsky2009}. These states are, in general, characterized by the reduced uncertainty in one field quadrature compared to the vacuum state, but having increased the uncertainty in the conjugate quadrature. Such states, with respect to light, were first generated in 1985 by Slusher et al. \cite{SHY85}, who achieved this through four-wave mixing in sodium atoms placed in an optical cavity. Shortly afterwards, in 1986, Shelby et al. also generated squeezed light using four-wave mixing in an optical fiber \cite{SLW86}. It has been also produced using degenerate parametric down-conversion in a second-order nonlinear crystal placed in an optical cavity \cite{WKH86}.

Similar to entangled states, squeezed states have a wide range of applications. As they are related to entanglement in multi-mode systems, they are a valuable resource \cite{DDA10,GH21,RPF22,GGS24,NDH25}. Squeezed states, for instance, are used in quantum cryptography protocols \cite{MUL12,GHD15,ABT24} and in quantum error correction \cite{HQ23,XZW23,KBG24}.

Three-qubit systems exhibit a rich structure that supports both bipartite and genuine tripartite entanglement. Examples of such states include the GHZ state and the W state, which represent distinct classes of multipartite entanglement~\cite{SG08}. It is crucial to understand how squeezing manifests in such three-qubit states and what is its relation to the entanglement considering different entanglement classes of the three-qubit system. Moreover, considering multi-qubit systems described by qubit-chain models or collective qubit interactions, the three-qubit states frequently act as elementary building blocks for larger entangled ensembles.

The structure of the paper is as follows. In Section~\ref{sec:states}, classification of three-qubit states is given. Section~\ref{sec:measures} provides definitions of the parameters of two- and three-mode entanglement and squeezing. Section~\ref{sec:results} is devoted to the analysis of the relations among the parameters quantifying two- and three-mode entanglement and squeezing considering different types of three-qubit states exhibiting full-tripartite entanglement. Our findings are summarized in the concluding Sec.~\ref{sec:conclusions}.

\section{\label{sec:states}Three-qubit states}

The analysis of squeezing in the three-qubit system is based on the classification of three-qubit entangled states proposed by Sab{\'i}n and Garc{\'i}a-Alcaine in~\cite{SG08}. This classification includes both pure and mixed states and divides the three-qubit states into three types:
\begin{itemize}
\item[I] - fully separable states,
\item[II] - states with only bipartite entanglement,
\item[III] - states with non-zero full tripartite entanglement.
\end{itemize}

\begin{figure}[ht]
\centering
\includegraphics[width=0.65\textwidth]{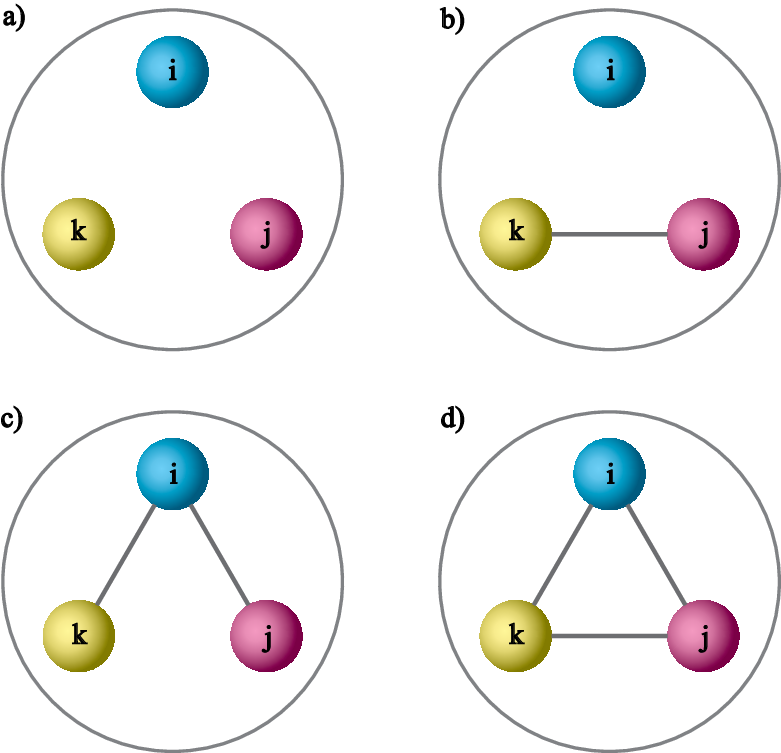}
\caption{Graphical representation of the states that are classified as type III: (a) subtype III-0, (b) subtype III-1, (c) subtype III-2, (d) subtype III-3.}\label{Fig1}
\end{figure}

We further concentrate on the entangled states exhibiting the full tripartite entanglement, i.e. belonging to type III. Graphical representation of the states classified as type III is presented in Fig.~\ref{Fig1}, where the qubits are represented by spheres labeled $i$, $j$, and $k$. The lines connecting these qubits represent two-mode entanglement whereas the outer circle denotes the full three-mode entanglement.
According to the classification of~\cite{SG08}, the states with non-zero full tripartite entanglement can be further divided into the following four subtypes:
\begin{itemize}
\item[III-0] - two-mode entanglement is missing, i.e. all reduced negativities are zero ($N_{ij}=N_{ik}=N_{jk}=0$) - see Fig.~\ref{Fig1}a,
\item[III-1] - two-mode entanglement between two qubits is observed, i.e. one reduced negativity is non-zero ($N_{ij}=N_{ik}=0$, $N_{jk}\neq 0$) - Fig.~\ref{Fig1}b,
\item[III-2] - two-mode entanglement inside two pairs of qubits is observed, i.e. two reduced negativities are non-zero ($N_{jk}=0$, $N_{ij}$ and $N_{ik}\neq 0$) - Fig.~\ref{Fig1}c,
\item[III-3] - two-mode entanglement inside all three pairs of qubits is observed, i.e. all three reduced negativities are non-zero ($N_{ij}$, $N_{ik}$, and $N_{jk}\neq 0$) - Fig.~\ref{Fig1}d.
\end{itemize}

\section{\label{sec:measures}Measures of squeezing and entanglement}
As a measure of entanglement of two modes, we use negativity. This measure is based on the negative-partial-transposition (NPT) criterion~\cite{P96,HHH96}. Negativity $ N_{ij} $ is defined as the sum of the absolute values of all negative eigenvalues $\lambda_{l}$ calculated for the two-qubit matrix $\rho_{ij}$ after partial transposition with respect to one of the modes. A two-mode density matrix $\rho_{ij}$ is derived from the three-mode density matrix $\rho_{ijk}$ by tracing out subsystem $k$
\begin{equation}
N_{ij}=-2 \sum_{l} \lambda_{l}\left(\rho_{ij}^{T_i}\right).
\end{equation}
We have $N_{ij}=0$ for separable states.

To quantify full tripartite entanglement, we use the geometric average of three negativities \cite{SG08,KCH22}
\begin{equation}
N_{ijk}=\left(N_{i-jk}N_{j-ik}N_{k-ij}\right)^{\frac{1}{3}},
\label{eq:Nijk}
\end{equation}
where $N_{i-jk}$, $N_{j-ik}$, and $N_{k-ij}$ are the bipartite negativities for the three-mode density matrix $\rho_{ijk}$, and the partial transposition is performed in turn for the modes $i$, $j$, and $k$. The negativity $N_{ijk}$ is always equal to zero for the states belonging to classes I and II. On the other hand, $N_{ijk}$ stands non-zero for the states of class III. For instance, assuming the state $\left(\vert 000 \rangle + \vert 111\rangle\right)/\sqrt{2}$ the negativity $N_{ijk}=1$. On the other hand, for the state $\left(\vert 001 \rangle + \vert 010 \rangle+\vert 100 \rangle\right)/\sqrt{3}$ the negativity $N_{ijk} =0.94$.

To reveal squeezing in two-mode states, we introduce the following two-mode quadrature operators:
\begin{eqnarray}
\hat{X}_{ij}&=&\hat{a}_i+\hat{a}_i^{\dagger}+\hat{a}_j+\hat{a}_j^{\dagger},\nonumber\\
\hat{Y}_{ij}&=&-i\left(\hat{a}_i-\hat{a}_i^{\dagger}+\hat{a}_j-\hat{a}_j^{\dagger}\right).
\label{3}
\end{eqnarray}
For these operators, the uncertainty relation takes the form:
\begin{equation}
 \langle\left(\Delta\hat{X}_{ij}\right)^2\rangle \langle\left(\Delta\hat{Y}_{ij}\right)^2\rangle \geq 4.
\end{equation}
Suitably rotating the field operators in the definitions (\ref{3}) we reveal the principal squeeze variance $ \lambda_{ij} $ that quantifies the maximal attainable squeezing of the field quadrature operators identified by the following inequality
~\cite{LPP88,KP97,AP00}:
\begin{eqnarray}
\lambda_{ij}=2\left[1+\langle\Delta\hat{a}_i^{\dagger}\Delta\hat{a}_i\rangle+\langle\Delta\hat{a}_j^{\dagger}\Delta\hat{a}_j\rangle+2Re\langle\Delta\hat{a}_i^{\dagger}\Delta\hat{a}_j\rangle \right. \nonumber \\
-\left. \vert\langle\left(\Delta\hat{a}_i\right)^2\rangle+\langle\left(\Delta\hat{a}_j\right)^2\rangle+2\langle\Delta\hat{a}_i\Delta\hat{a}_j\rangle \vert \right]<2
\label{eq:lambda_ij}
\end{eqnarray}
and $\langle\Delta\hat{a}_i^{\dagger}\Delta\hat{a}_i\rangle =
\langle\hat{a}_i^{\dagger}\hat{a}_i\rangle-\langle\hat{a}_i^{\dagger}\rangle\langle\hat{a}_i\rangle$.

Similarly, we define the three-mode quadrature operators $\hat{X}_{ijk} $ and $ \hat{Y}_{ijk} $ to judge three-mode squeezing:
\begin{eqnarray}
\hat{X}_{ijk}&=&\hat{a}_i+\hat{a}_i^{\dagger}+\hat{a}_j+\hat{a}_j^{\dagger}+\hat{a}_k+\hat{a}_k^{\dagger},\nonumber\\
\hat{Y}_{ijk}&=&-i\left(\hat{a}_i-\hat{a}_i^{\dagger}+\hat{a}_j-\hat{a}_j^{\dagger}+\hat{a}_k-\hat{a}_k^{\dagger}\right).
\end{eqnarray}
The corresponding three-mode principal squeeze variance satisfies the following inequality:
\begin{eqnarray}
    \lambda_{ijk}&=&3+2\left(\langle\Delta\hat{a}_i^{\dagger}\Delta\hat{a}_i\rangle+\langle\Delta\hat{a}_j^{\dagger}\Delta\hat{a}_j\rangle+\langle\Delta\hat{a}_k^{\dagger}\Delta\hat{a}_k\rangle\right) \nonumber \\
    &&+4Re\left\{\langle\Delta\hat{a}_i^{\dagger}\Delta\hat{a}_j\rangle+\langle\Delta\hat{a}_i^{\dagger}\Delta\hat{a}_k\rangle+\langle\Delta\hat{a}_j^{\dagger}\Delta\hat{a}_k\rangle \right\} \nonumber \\
    &&-2 \vert\langle\left(\Delta\hat{a}_i\right)^2\rangle+\langle\left(\Delta\hat{a}_j\right)^2\rangle+\langle\left(\Delta\hat{a}_k\right)^2\rangle+2\left(\langle\Delta\hat{a}_i\Delta\hat{a}_j\rangle\right. \nonumber\\
    &&\left.+\langle\Delta\hat{a}_i\Delta\hat{a}_k\rangle+\langle\Delta\hat{a}_j\Delta\hat{a}_k\rangle\right) \vert <3 ,
    \label{eq:lambda_ijk}
\end{eqnarray}
We note that, in this case, the quadrature uncertainty relation takes the form:
\begin{equation}
 \langle\left(\Delta\hat{X}_{ijk}\right)^2\rangle\langle\left(\Delta\hat{Y}_{ijk}\right)^2\rangle \geq 9.
\end{equation}

In continuous-variable (CV) systems, squeezing refers to the situation in which the noise in one quadrature of an optical field is reduced below the standard quantum limit. This is accompanied by the increases of the noise in the conjugated quadrature. Though the description of CV systems reached by using the quantum scissors \cite{LK11} is restricted to few states and its dynamics resembles that of discrete systems with couple of quantum levels, physical meaning of squeezing remains the same.

We note that, for qubits, the general concept of squeezing results in the spin squeezing. In these systems, that can be considered as spin $1/2$ particles, squeezing refers to reducing the uncertainty (variance) in one component of the collective spin operator below the standard quantum limit. This is achieved at the expense of increased uncertainty in another complementary component and enables enhanced precision in estimating small rotations around an orthogonal axis. This is analogous to standard CV quadrature squeezing. Spin squeezing arises naturally in ensembles of qubits (e.g., atoms or ions) and, similarly as in the analyzed system, it can indicate the presence of non-classical correlations \cite{KCL05,MWS11}.

Simplicity of the projection measurements in qubit systems motivate us to study squeezing in truncated bosonic systems where it can be a useful resource for 'qubit-based' quantum information processing. The analyzed qubit-squeezed states and the spin-squeezed states in particular have a number of practical applications in atomic clocks and quantum sensors. They help to overcome the limitations imposed by the projection noise. For small cold-atom ensembles and related metrological platforms, these states can significantly increase precision \cite{G12}.

Moreover, the truncated bosonic systems serve as a source of strong quantum correlations, including entanglement, quantum steering and Bell nonlocality \cite{KKJ19,KK19}. Nair and Flebus demonstrated in \cite{NF25} that spin squeezing can form multiparty entanglement, which is a crucial resource in quantum metrology and information processing.

We note that the standard quantum limit is derived from the Heisenberg uncertainty principle and equals to 2 (3) for two- (three-) mode variances.

\section{\label{sec:results}Results}
Considering the states exhibiting tripartite entanglement ($N_{ijk} > 0 $) and using the above division of these states into sub-classes III-0, III-1, III-2, and III-3, we systematically examine the two- and three-mode squeezing and their relation to two- and three-mode negativity.

\subsection{\label{sec:III_0}Type III-0}

For the states belonging to type III-0, all two-mode negativities are zero ($N_{ij}=N_{ik}=N_{jk}= 0$) (see Figure~\ref{Fig1}a) and the states are parameterized as follows:
\begin{equation}
\vert\psi\rangle=C_{000}\vert 000 \rangle +C_{111}\vert 111\rangle
\label{eq:psi_III-0}
\end{equation}
where $C_{000}$ and $C_{111}$ are the probability amplitudes of the states $\vert 000 \rangle$ and $\vert 111\rangle$.

In our numerical investigations, we have randomly generated the three-qubit states described by the density matrix $\rho\equiv\vert \psi \rangle\langle\psi\vert$ approximately $\sim10^4$ times. In detail, we generated $\sim10^4$ pairs of numbers in which the first number is $C_{000}$ and the second one $C_{111}$. These numbers satisfy the condition $C_{000}C_{000}^*+C_{111}C_{111}^*=1$. Then we created a state $\psi$ according to the formula (\ref{eq:psi_III-0}) and calculated the corresponding density matrix $\rho$ for each pair. Finally, the negativities and the principal squeeze variances were determined for each realization.

\begin{figure}[ht]
\centering
\includegraphics[width=0.42\textwidth]{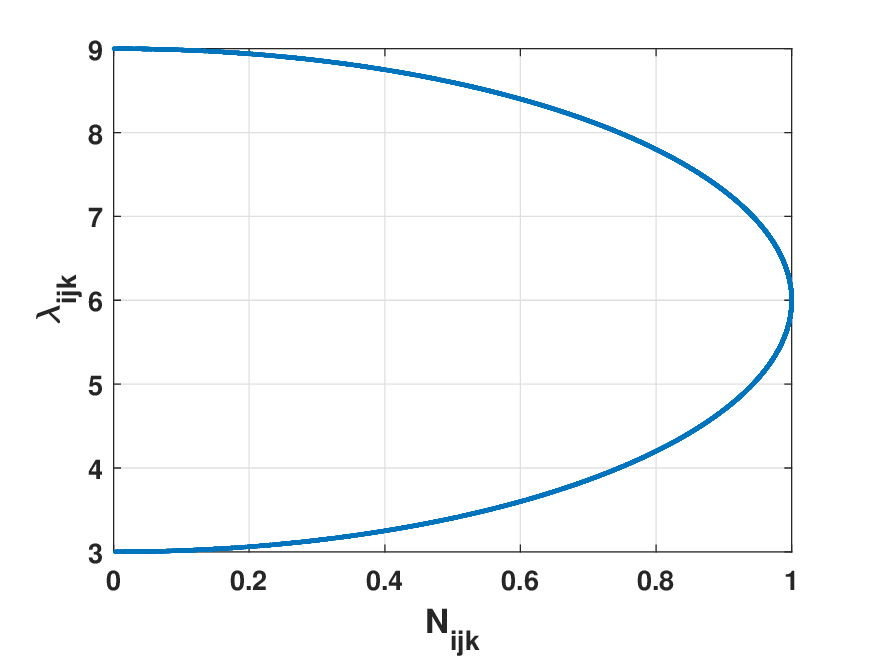}
\caption{\label{Fig_III-0} Three-mode principal squeeze variance $\lambda_{ijk} $ as it depends on three-mode negativity $ N_{ijk}$ for the states of the type III-0.}
\end{figure}

We express the two- and three-mode principal squeeze variances for the system described by the wave function given in equation (\ref{eq:psi_III-0}) using formulas (\ref{eq:lambda_ij}) and (\ref{eq:lambda_ijk}) and the probability $P_{111}= |C_{111}|^2$ as follows:
\begin{eqnarray}
\lambda_{ij}&=&\lambda_{jk}=\lambda_{ik}=2+4P_{111},\nonumber \\
\lambda_{ijk}&=&3+6P_{111}.
\label{eq:lambda_III_0}
\end{eqnarray}
For simplicity, we assume that the probability amplitudes are real. This assumption is used in this and the subsequent analysis. It reveals that the two-mode principal squeeze variance for each pair of qubits never reaches the values smaller than $2$ and  the three-mode principal squeeze variance is always equal to or greater than $3$. Therefore, two- and three-mode squeezing does not occur for states belonging to subtype III-0.  We note that the symmetry of the states ensures that $ \lambda_{ij} = \lambda_{ik} =\lambda_{jk} $ and $ N_{ij} = N_{ik} = N_{jk} =0 $. Interestingly, the three-mode negativity $ N_{ijk} $ and three-mode principal squeeze variance $ \lambda_{ijk} $ form a specific relation described in the graph of Fig.~\ref{Fig_III-0}.

\subsection{\label{sec:III_1}Type III-1}

The states belonging to type III-1 ($N_{ij}=N_{ik}=0$, $N_{jk}\neq0$, see Fig.~\ref{Fig1}b) can be divided into the following two subtypes:
\begin{itemize}
\item[.] the subtype III-1A
\begin{equation}
\vert\psi\rangle=C_{000}\vert 000\rangle +C_{\vert 1\rangle_{i}\vert 00\rangle_{jk}}\vert 1\rangle_{i} \otimes \vert 00\rangle_{jk} +C_{111}\vert 111\rangle
\label{eq:psi_III-1A}
\end{equation}
\item[.] the subtype III-1B
\begin{equation}
\vert\psi\rangle=C_{000}\vert 000\rangle +C_{\vert0\rangle_{i}\vert 11\rangle_{jk}}\vert 0\rangle_{i} \otimes\vert11\rangle_{jk}+C_{111}\vert 111\rangle. \label{eq:psi_III-1B}
\end{equation}
\end{itemize}
To introduce the notation used in Eqs.~(\ref{eq:psi_III-1A}) and (\ref{eq:psi_III-1B}), the states belonging to subtype III-1A are explicitly expressed as $C_{000}\vert 000 \rangle +C_{100}\vert100 \rangle +C_{111}\vert 111\rangle$, $C_{000}\vert 000 \rangle +C_{010}\vert 010 \rangle +C_{111}\vert 111\rangle$, and $C_{000}\vert 000 \rangle +C_{001}\vert 001 \rangle +C_{111}\vert 111\rangle$.

We note that the states of subtype III-1A can be mapped to the states of subtype III-1B by mutual replacing the states $\vert0\rangle $ and $ \vert1\rangle $ in each mode. Nevertheless, as these states belong to the states of quantum harmonic oscillator, the corresponding states of subtype III-1A and III-1b exhibit different squeezing properties.

To reveal the properties of the states given in Eqs.~(\ref{eq:psi_III-1A}) and (\ref{eq:psi_III-1B}), we have randomly generated $\sim 10^5$ states. In the following figures, in which we analyze the behavior of two- and three-mode squeezing, yellow regions denote the occurrence of states exhibiting squeezing, i.e. $\lambda_{jk}<2$ or $\lambda_{ijk}<3$. The regions encompassing the remaining states are drawn in green. To achieve better understanding, in these figures we also draw the curves belonging to the states in which one of the three probabilities equals zero. The black, red, and blue curves belong in turn to the states in which $P_{000}=|C_{000}|^2=0$, $P_{111}=0$, or $P_{\vert 1\rangle_{i}\vert 00\rangle_{jk}}=0$.

\begin{figure*}[ht]
\centering
\includegraphics[width=0.42\textwidth]{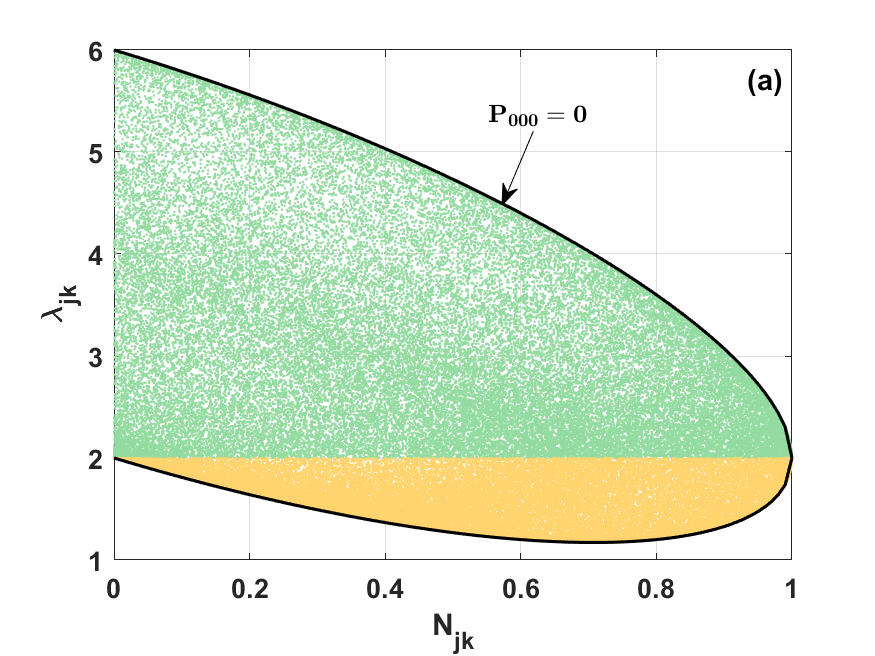}
\includegraphics[width=0.42\textwidth]{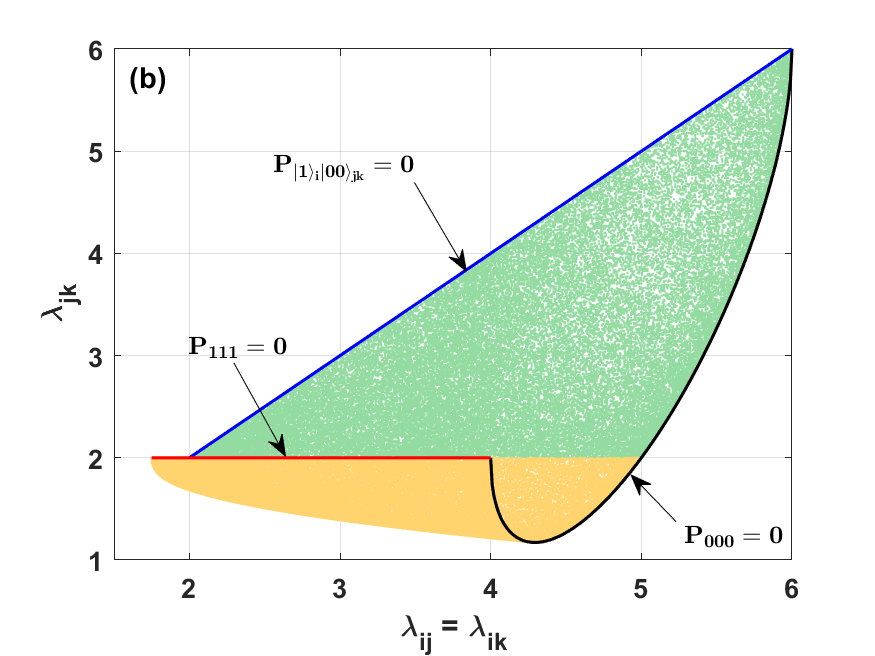}\\
\includegraphics[width=0.42\textwidth]{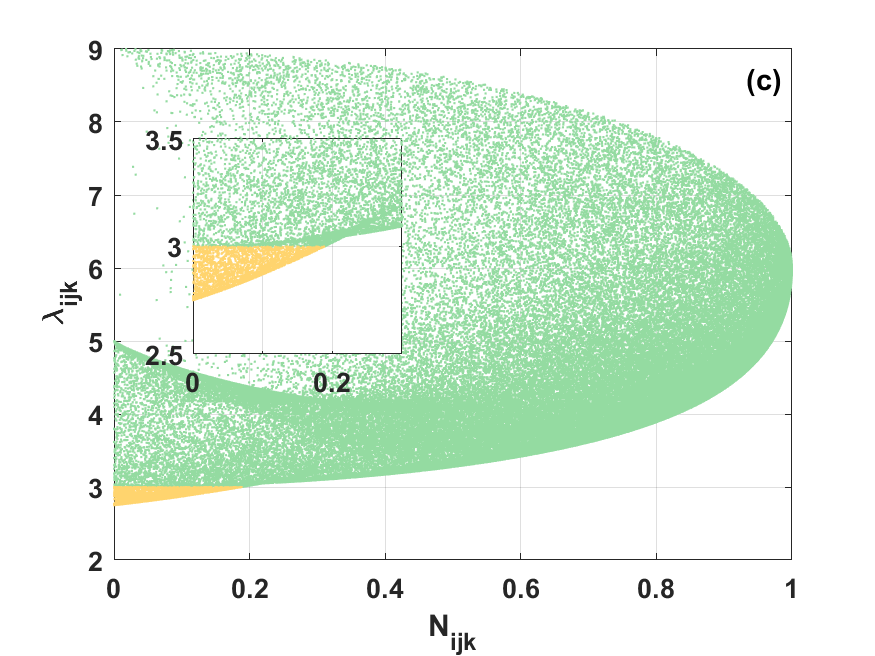}\\
\includegraphics[width=0.42\textwidth]{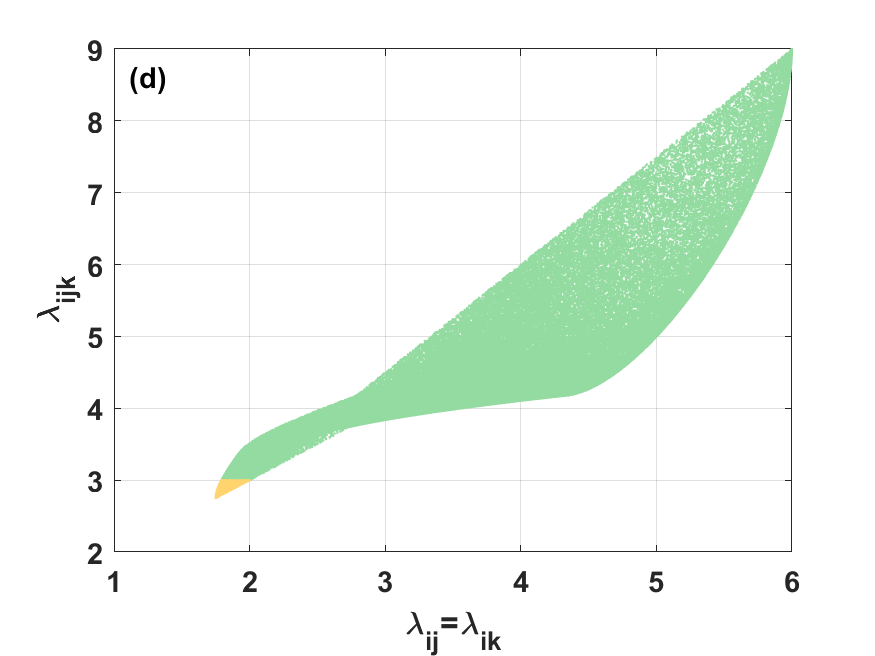}
\includegraphics[width=0.42\textwidth]{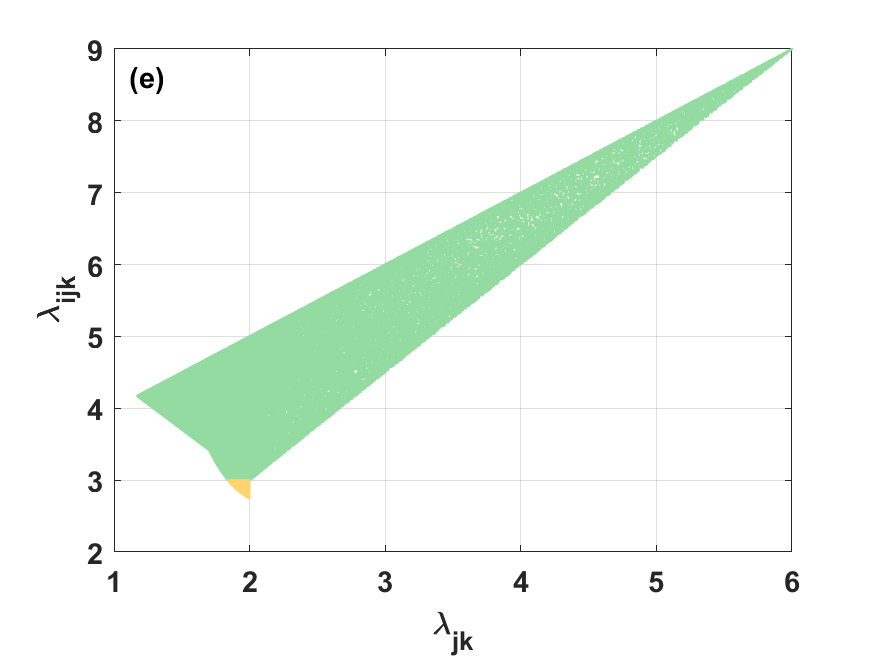}
\caption{\label{Fig_III-1A} Graphs revealing the relations among two- and three-mode principal squeeze variances $\lambda_{jk}$, $\lambda_{ij} = \lambda_{ik} $, and $\lambda_{ijk} $ and two- and three-mode negativities $ N_{jk}$ and $N_{ijk} $ for the states of the type III-1A.}
\end{figure*}

Considering the states described by the wave function
\begin{equation}
\vert\psi\rangle=C_{000}\vert 000 \rangle +C_{100}\vert100 \rangle +C_{111}\vert 111\rangle,
\label{eq:psi_000_100_111}
\end{equation}
the two- and three-mode principal squeeze variances are defined as:
\begin{eqnarray}
\lambda_{ij}&=&\lambda_{ik}=2\left( 1+P_{100}+2P_{111}-2P_{000}P_{100}\right), \nonumber \\
\lambda_{jk}&=&2\left( 1+2P_{111}-2\sqrt{P_{100}P_{111}}\right), \nonumber \\
\lambda_{ijk}&=&3+2\left(P_{100}-P_{000}P_{100}+3P_{111} \right)-2\mid 2\sqrt{P_{100}P_{111}}-P_{000}P_{100}\mid.
\label{eq:lambda_III_1A}
\end{eqnarray}
Analyzing these formulas, we can see that the principal squeeze variances for two and three modes can be lower than $2$ and $3$, respectively. Therefore, two- and three-mode squeezing occurs for some states characterized by Eq~(\ref{eq:psi_000_100_111}). A more detailed analysis of the states classified as subtype III-1A is given in Fig.~\ref{Fig_III-1A}.

In Fig.~\ref{Fig_III-1A}a we analyze the relation between the two-mode squeezing $ \lambda_{jk} $ and the only two-mode nonzero negativity $N_{jk}$. The black solid curve gives the minimal and maximal border values of the squeezing parameter $\lambda_{jk}$. Whereas the minimal values of $\lambda_{jk}$ are obtained for the probabilities $P_{000} = 0$ and $P_{111} \in \langle 0; 0.5\rangle $, the maximal values occur when $P_{000}=0$ and $P_{111} \in
\langle 0.5; 1\rangle $. In this specific case, the wave function (\ref{eq:psi_III-1A}) simplifies into the form $\vert\psi\rangle=C_{\vert 1\rangle_{i}\vert 00\rangle_{jk}}\vert 1\rangle_{i} \otimes \vert 00\rangle_{jk} +C_{111}\vert 111\rangle$. According to the curves in Fig.~\ref{Fig_III-1A}a, the smallest value of the principal squeeze variance $\lambda_{jk}\approx 1.17$ occurs for the negativity $N_{jk}\approx 0.71$ and the probability $P_{111}=\frac{1}{4} \left(2-\sqrt{2}\right)\approx 0.15$. We note that, at this point, $\lambda_{ij}=\lambda_{ik}=2 \left(\frac{1}{4} \left(2-\sqrt{2}\right)+2\right)\approx 4.29$ (see the black curve in Fig.~\ref{Fig_III-1A}b, and $\lambda_{ijk}=-2 \sqrt{\left(2-\sqrt{2}\right) \left(\frac{1}{4} \left(\sqrt{2}-2\right)+1\right)}-\sqrt{2}+7\approx 4.17$ indicates no other form of squeezing in the analyzed state.

Nevertheless, two-mode squeezing in modes $i-j$ and $i-k$ is also reached, as documented in Fig.~\ref{Fig_III-1A}b. In this case, the principal squeeze variances $\lambda_{ij}=\lambda_{ik}=1.75$, $\lambda_{jk}=2$ and the corresponding states attain the form $\vert\psi\rangle=C_{000}\vert 000\rangle +C_{\vert 1\rangle_{i}\vert 00\rangle_{jk}}\vert 1\rangle_{i} \otimes \vert 00\rangle_{jk} $.

The relation between three-mode squeezing and tripartite entanglement is illustrated in Fig.~\ref{Fig_III-1A}c. Only a small number of states exhibit tripartite squeezing, specifically those with rather weak tripartite entanglement. The three-mode negativity $N_{ijk}$ cannot exceed $\approx 0.19$ to observe the three-mode squeezing. The smallest value of three-mode principal squeeze variance $\lambda_{ijk}= 2.75$ occurs for $N_{ijk} = 0$. The two-mode principal squeeze variances equal $\lambda_{jk}=2$ and $\lambda_{ij}=\lambda_{ik}=1.75$ at this point.

Figures~\ref{Fig_III-1A}d and~\ref{Fig_III-1A}e illustrate the relation between two- and three-mode squeezing. In the yellow area the states exhibiting three-mode squeezing ($\lambda_{ijk}<3$) are found. The strongest three-mode squeezing occurs when the parameters $\lambda_{ij}$ and $\lambda_{ik}$ reach their minimal value. The strongest two-mode squeezing also occurs for two pairs of qubits. However, the remaining pair of qubits, $j-k$, is not squeezed ($\lambda_{jk}\approx 2$). Importantly, three-mode squeezing occurs when at least two of the three two-mode squeezing parameters are smaller than $2$.

Now let us analyze the states classified as subtype III-1B and expressed as $C_{000}\vert 000 \rangle +C_{011}\vert 011 \rangle +C_{111}\vert 111\rangle$, $C_{000}\vert 000 \rangle +C_{101}\vert 101 \rangle +C_{111}\vert 111\rangle$, and $C_{000}\vert 000 \rangle +C_{110}\vert 110 \rangle +C_{111}\vert 111\rangle$.

\begin{figure*}[ht]
\centering
\includegraphics[width=0.42\textwidth]{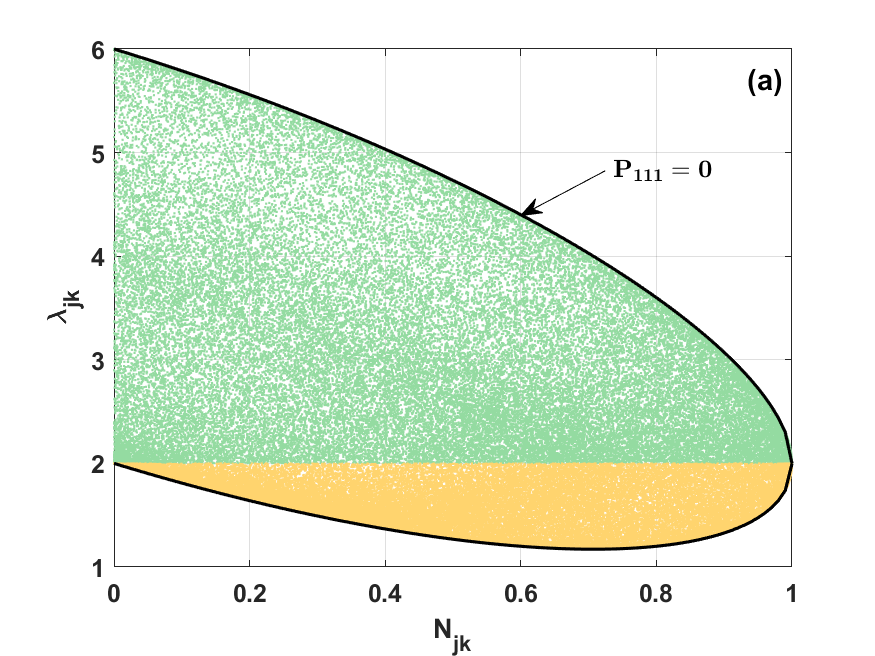}
\includegraphics[width=0.42\textwidth]{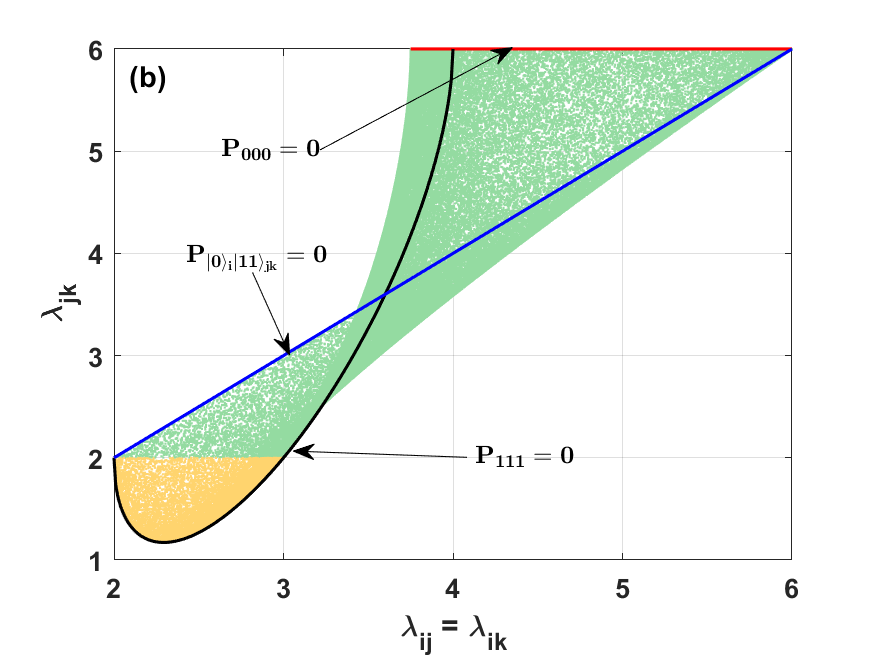}\\
\includegraphics[width=0.42\textwidth]{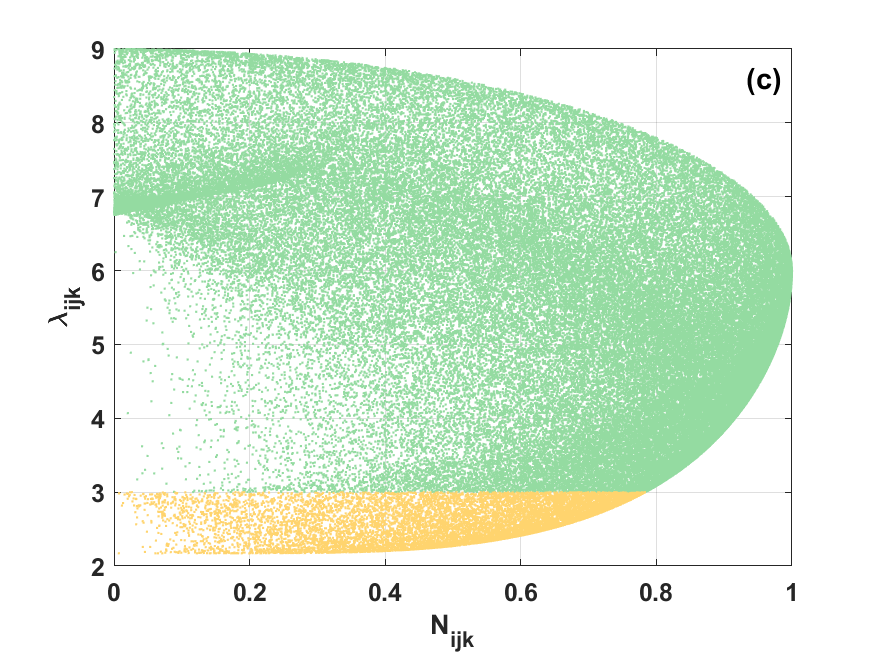}\\
\includegraphics[width=0.42\textwidth]{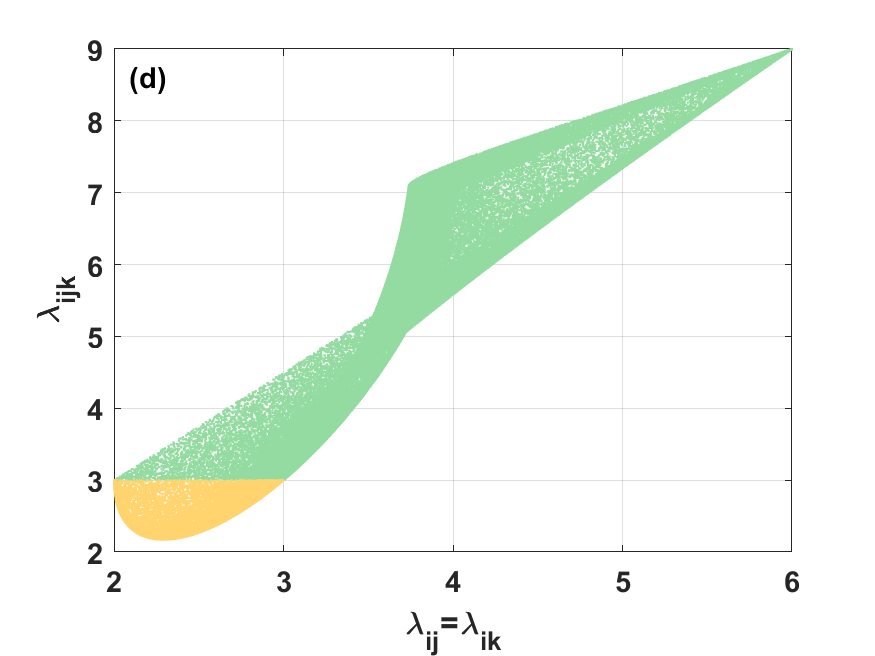}
\includegraphics[width=0.42\textwidth]{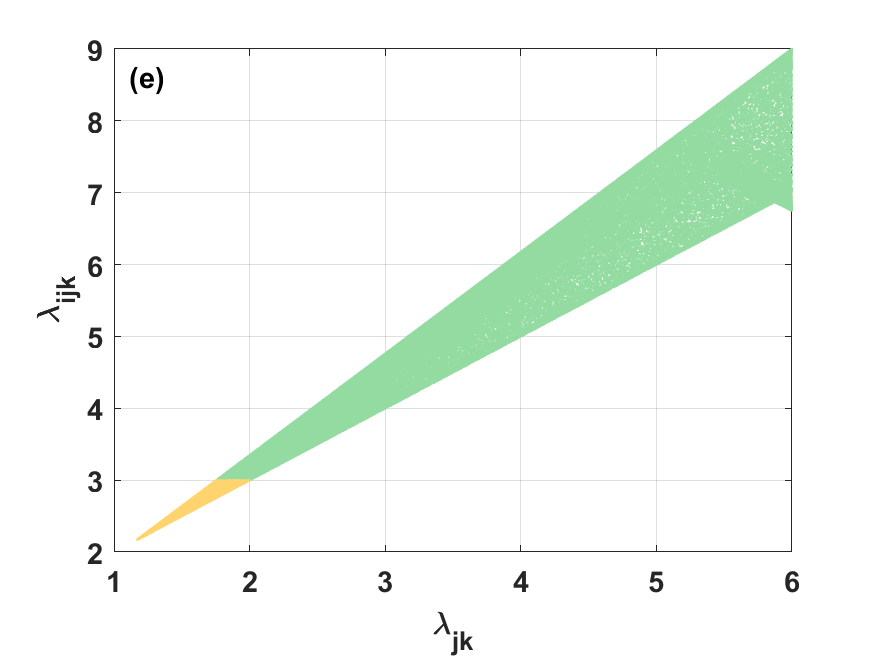}
\caption{\label{Fig_III-1B} Graphs revealing the relations among two- and three-mode principal squeeze variances $\lambda_{jk}$, $\lambda_{ij} = \lambda_{ik} $, and $\lambda_{ijk} $ and two- and three-mode negativities $ N_{jk}$ and $N_{ijk} $ for the states of the type III-1B.}
\end{figure*}

Expressing the principal squeeze variances by the probabilities for the state
\begin{equation}
\vert\psi\rangle=C_{000}\vert 000 \rangle +C_{011}\vert011 \rangle +C_{111}\vert 111\rangle,
\label{eq:psi_000_011_111}
\end{equation}
we reveal the following formulas:
\begin{eqnarray}
\lambda_{ij}&=&\lambda_{ik}=2\left( 1+P_{011}+2P_{111}-2P_{011}P_{111}\right),\nonumber \\
\lambda_{jk}&=&2\left( 1+2P_{011}+2P_{111}-2\sqrt{P_{000}P_{011} }\right),\nonumber \\
\lambda_{ijk}&=&3+2\left(P_{011}-P_{011}P_{111}+3P_{111} \right)-2\mid 2\sqrt{P_{000}P_{011}}-P_{011}P_{111}\mid.
\label{eq:lambda_III_1B}
\end{eqnarray}
According to these formulas, for a given set of probabilities satisfying the relation $P_{000}+P_{011}+P_{111}=1$ the variances $\lambda_{jk}$ and $\lambda_{ijk}$ are lower than $2$ and $3$, respectively. Thus, some of the states characterized by the wave function (\ref{eq:psi_000_011_111}) are squeezed. Further analysis of states classified as subtype III-1B is presented in Fig.~\ref{Fig_III-1B}.

The relation between the two-mode principal squeeze variance $\lambda_{jk}$ and bipartite negativity $ N_{jk} $ is elucidated in Fig.~\ref{Fig_III-1B}a. Contrary to the similar graph in Fig.~\ref{Fig_III-1A}a, the black solid curve in Fig.~\ref{Fig_III-1B}a giving the minimal [maximal] values of the principal squeeze variance $\lambda_{jk}$ is obtained for $P_{111} = 0 $ and $ P_{000} \in\langle 0.5; 1\rangle $ [$P_{111} = 0 $ and $ P_{000} \in\langle 0; 0.5\rangle $]. In accordance with the graph in Fig.~\ref{Fig_III-1A}a, the smallest value of the principal squeeze variance $\lambda_{jk}\approx 1.17$ is reached for $N_{jk}\approx 0.71$ and $P_{000}=\frac{1}{4} \left(2+\sqrt{2}\right)\approx 0.85$.

In contrast to the III-1A states, the III-1B states do not exhibit two-mode squeezing quantified by $ \lambda_{ij} = \lambda_{ik}$, as shown in Fig.~\ref{Fig_III-1B}b. On the other hand, the state that minimizes the two-mode principal squeeze variance $\lambda_{jk} $ also minimizes the three-mode principal squeeze variance $\lambda_{ijk} $ at the value $\lambda_{ijk}=-2 \sqrt{\left(2+\sqrt{2}\right) \left(\frac{1}{4} \left(-2-\sqrt{2}\right)+1\right)}-\sqrt{2}+5\approx 2.17$, and gives $\lambda_{ij}=\lambda_{ik}=2 \left(\frac{1}{4} \left(-2-\sqrt{2}\right)+2\right)\approx 2.29$ (compare Figs.~\ref{Fig_III-1B}d and~\ref{Fig_III-1B}e).

Following the graph in Fig.~\ref{Fig_III-1B}c the range of negativities $N_{ijk}$ allowing three-mode squeezing is much broader ($ N_{ijk} < 0.78$) compared to that for the states of subtype III-1A. Also here, the largest squeezing occurs for $N_{ijk}= 0 $ when the principal squeeze variance $ \lambda_{ijk} \approx 2.17$. Nevertheless, for the states with $ N_{ijk} < \approx 0.3$, the values of $\lambda_{ijk}$ are close to the minimal achievable value $2.17$.

Figures~\ref{Fig_III-1B}d and~\ref{Fig_III-1B}e quantify the relationship between the two- and three-mode squeezing. The states exhibiting three-mode squeezing occupy the yellow region. For the states classified as III-1B both three-mode squeezing and squeezing of the $j-k$ qubit pair appear simultaneously. This resembles the situation in III-1A states where the strongest three-mode squeezing occurs simultaneously with the strongest two-mode squeezing. Additionally, three-mode squeezing can occur when no squeezing for the pairs of qubits $i-j$ and $i-k$ is found. In this case the parameters $\lambda_{ij}$ and $\lambda_{ik}$ take on values in the range $\left<2;3\right>$. Provided that values of these two parameters exceed $3$  three-mode squeezing does not occur.

\begin{figure}[ht]
\centering
\includegraphics[width=0.42\textwidth]{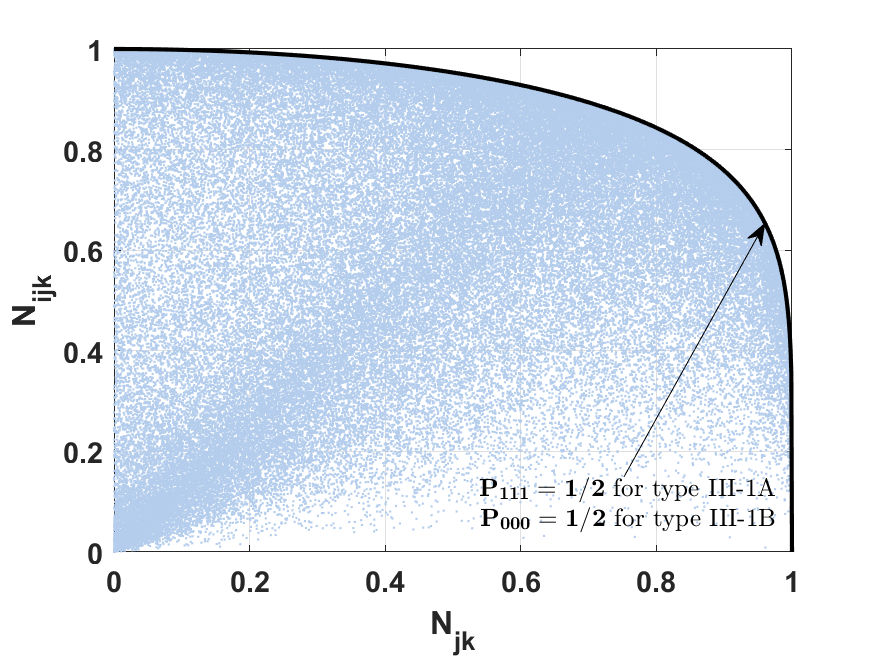}
\caption{\label{Fig_III-1neg} Graphs revealing the relation between three- and two-mode negativities $ N_{ijk}$ and $N_{jk} $ for the states of the type III-1A and III-1B.}
\end{figure}

Although the states of subtypes III-1A and III-1B exhibit different squeezing properties they exhibit similar entanglement properties. Figure~\ref{Fig_III-1neg} shows the relation between two- and three-mode negativities. The maximal three-mode entanglement is obtained when the entanglement between qubits $j$ and $k$ disappears. This occurs when the probability $P_{000}$ or $P_{111}$ is zero for the states of type III-1A and III-1B, respectively. This means that the maximal three-mode entanglement is achieved when the two-mode negativities are equal ($N_{ij}=N_{jk}=N_{ik}=0$). In contrast, as the entanglement between qubits $j$ and $k$ increases, the maximal attained value of the three-mode negativity decreases. At the boundary, the three-mode entanglement disappears and the two-mode entanglement is maximal ($N_{jk}=1$). Importantly, the maximal value of $ N_{ijk}$ for a given $N_{jk} $ is obtained when $P_{111}=1/2$ or $P_{000}=1/2$ for subtypes III-1A and III-1B, respectively.

\subsection{\label{sec:III_2}Type III-2}

Apart from the full tripartite entanglement, the states classified as type III-2 (see Figure~\ref{Fig1}c) exhibit two-mode entanglement inside two pairs of qubits ($N_{ij} > 0$, $N_{ik}>0$, $N_{jk}=0$). These states are parameterized as follows:
\begin{equation}
\vert\psi\rangle=C_{000}\vert 000\rangle +C_{\vert 1\rangle_{k}\vert 00\rangle_{ij}}\vert 1\rangle_{k} \otimes\vert 00\rangle_{ij}
+C_{\vert 0\rangle_{j}\vert 11\rangle_{ik}}\vert 0\rangle_{j} \otimes\vert 11\rangle_{ik}+C_{111}\vert 111\rangle
\label{eq:psi_III-2}
\end{equation}
They include the states $C_{000}\vert 000 \rangle +C_{001}\vert 001 \rangle +C_{011}\vert 011 \rangle+C_{111}\vert 111\rangle$, $C_{000}\vert 000 \rangle +C_{010}\vert 010 \rangle+C_{110}\vert 110 \rangle+C_{111}\vert 111\rangle$, and $C_{000}\vert 000 \rangle +C_{100}\vert 100 \rangle+C_{101}\vert 101 \rangle +C_{111}\vert 111\rangle$.

\begin{figure}[ht]
\centering
\includegraphics[width=0.42\textwidth]{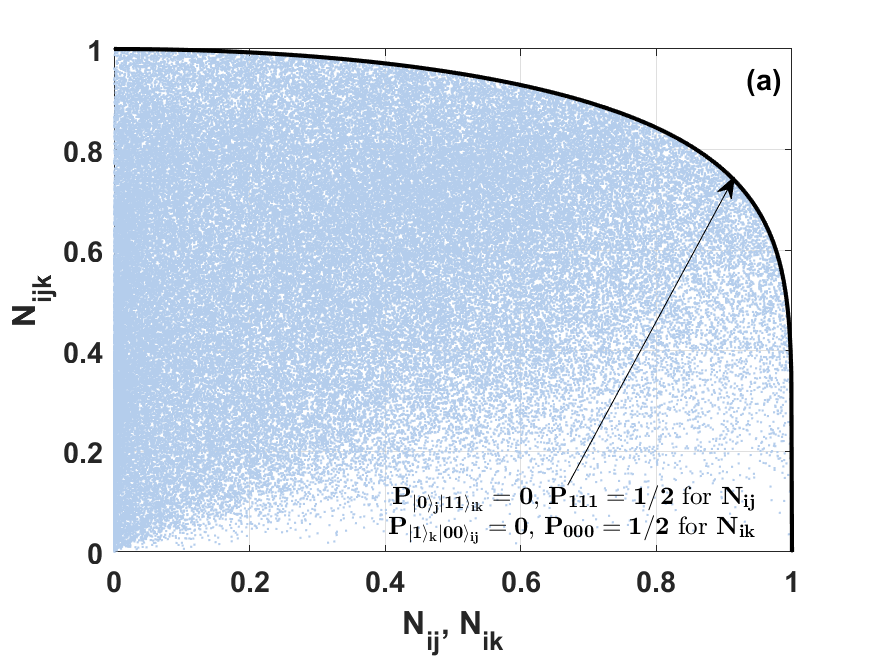}
\includegraphics[width=0.42\textwidth]{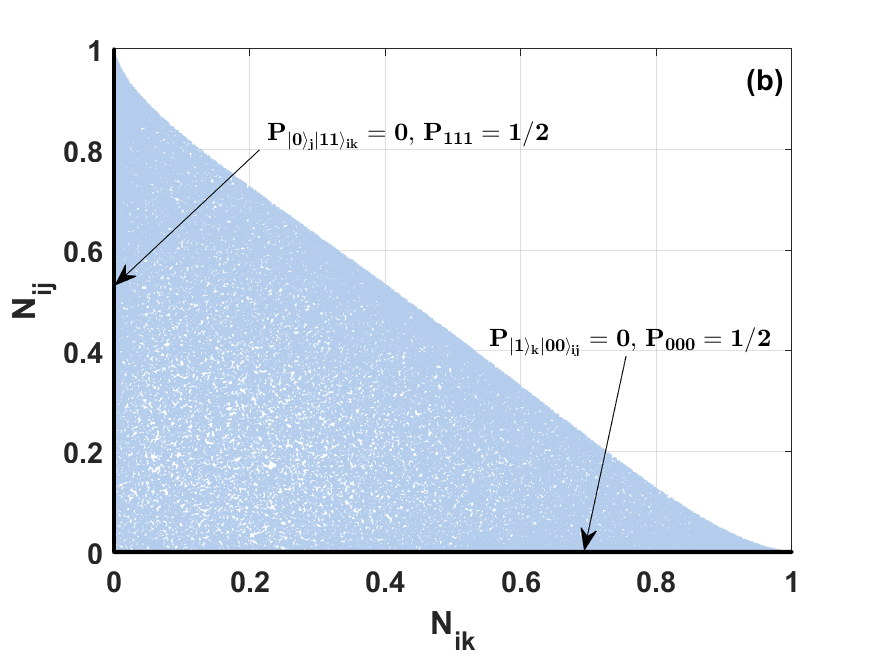}
\caption{\label{Fig_III-2neg} Graphs revealing the relation among three- and two-mode negativities $ N_{ijk}$, $N_{ij} $ and $N_{ik} $ for the states of the type III-2.}
\end{figure}

Even though the states described by the equation~(\ref{eq:psi_III-2}) belong to type III-2, their relation between two- and three-mode negativities is the same as for the III-1A and III-1B states (see Fig.~\ref{Fig_III-2neg}a). The maximal value of three-mode negativity $N_{ijk}$ for a given two-mode negativity $N_{ij}$ or $N_{ik}$ is obtained when $P_{|0\rangle_j|11\rangle_{ik}}=0$, $P_{111}=1/2$ or $P_{|1\rangle_k|00\rangle_{ij}}=0$, $P_{000}=1/2$, respectively. Comparing Figs.~\ref{Fig_III-2neg}a and \ref{Fig_III-2neg}b, the maximal value of $N_{ijk}$ for a given $N_{ij}$ is obtained when $N_{ik}=0$, and vice versa. This means that the maximal three-mode entanglement is achieved when all two-mode negativities are equal to zero. On the other hand, when the two-mode entanglement for one of the qubit pairs $i-j$ or $i-k$ reaches its maximal value, the three-mode entanglement disappears. Importantly, if one of the two-mode negativities reaches its maximal value, then the other has to be equal to zero (if $N_{ij} = 1$ then $N_{ik}=0$; if $N_{ik}=1$ then $N_{ij}=0$).

\begin{figure}[ht]
\centering
\includegraphics[width=0.42\textwidth]{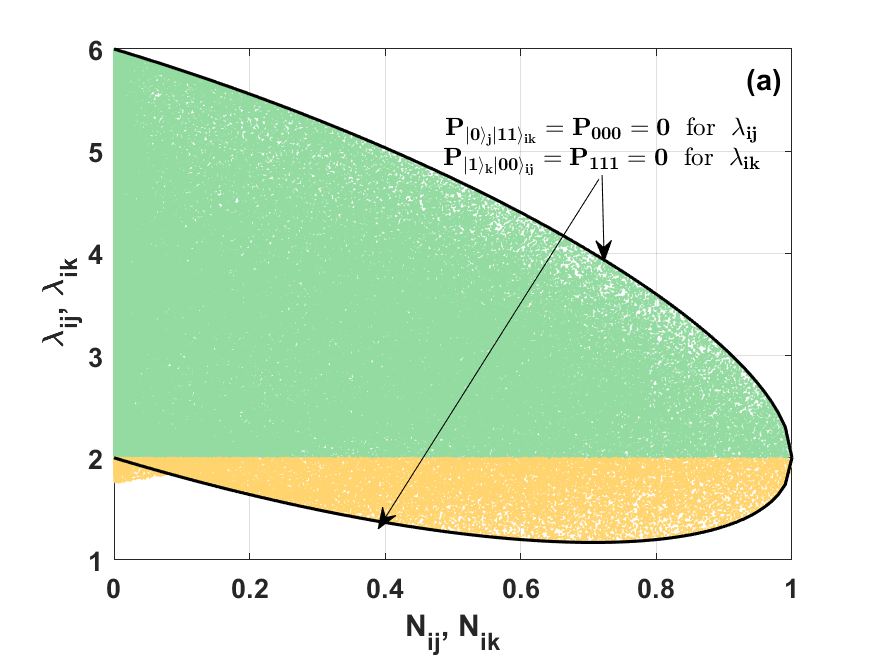}
\includegraphics[width=0.42\textwidth]{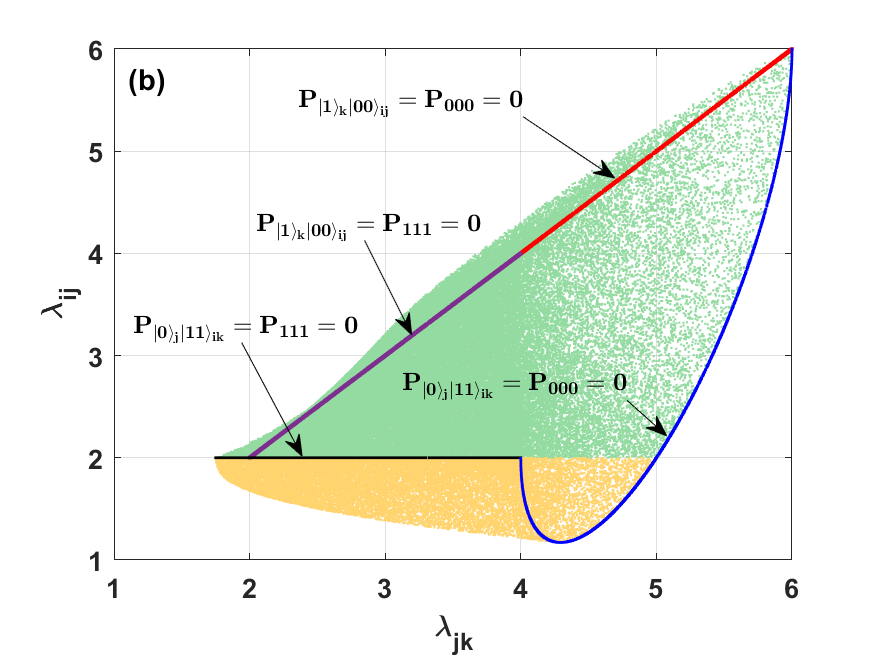}\\
\includegraphics[width=0.42\textwidth]{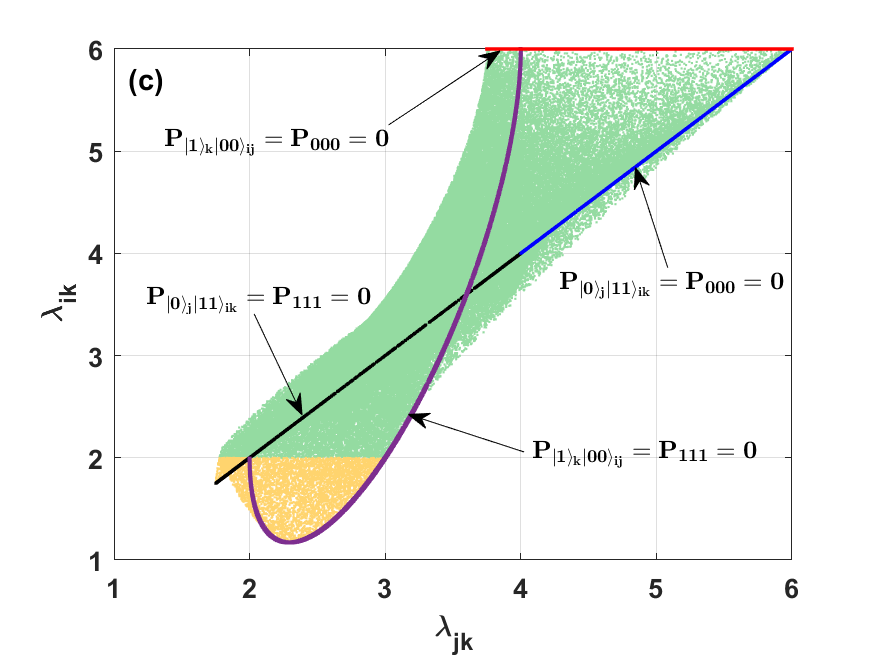}
\includegraphics[width=0.42\textwidth]{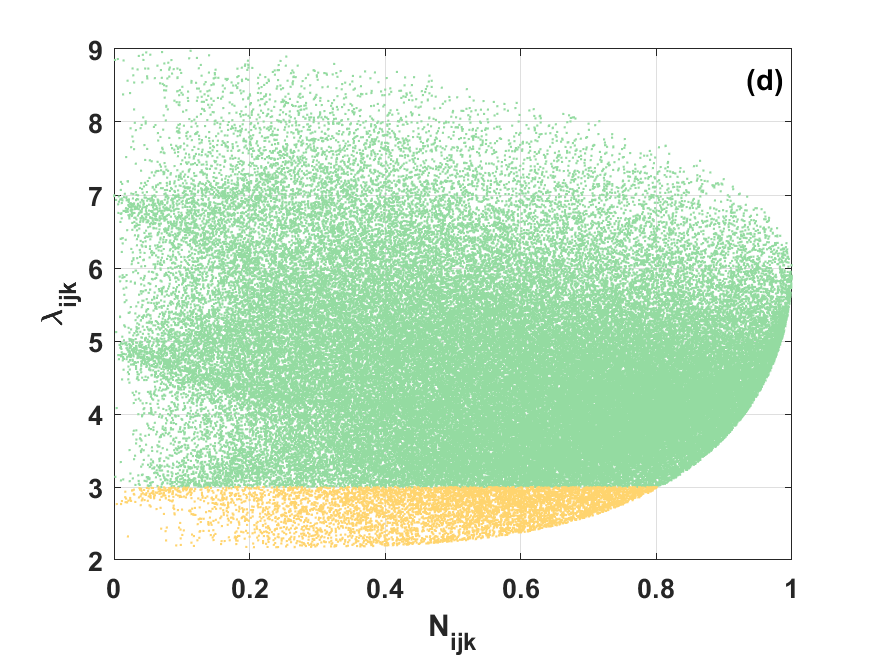}\\
\includegraphics[width=0.42\textwidth]{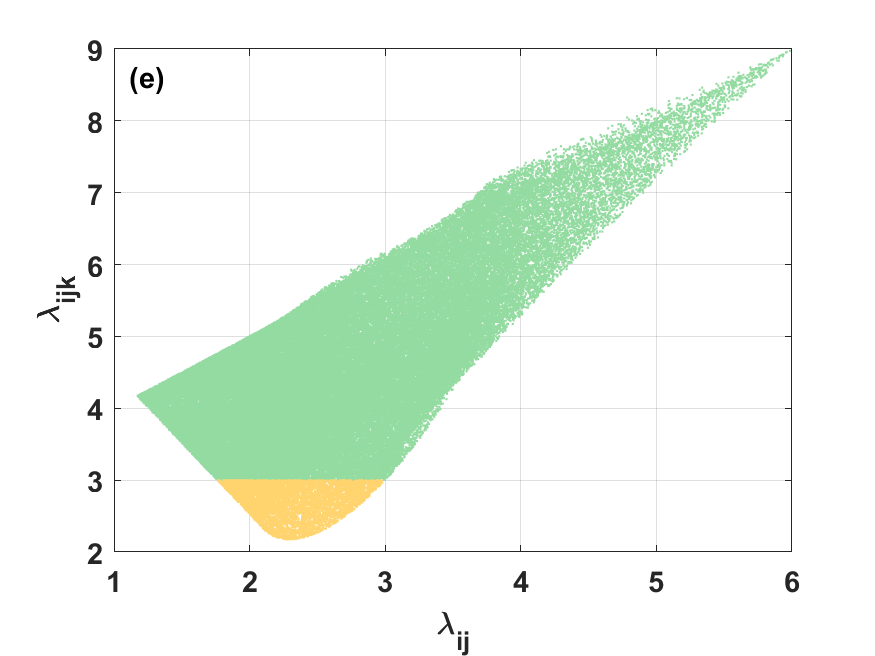}
\includegraphics[width=0.42\textwidth]{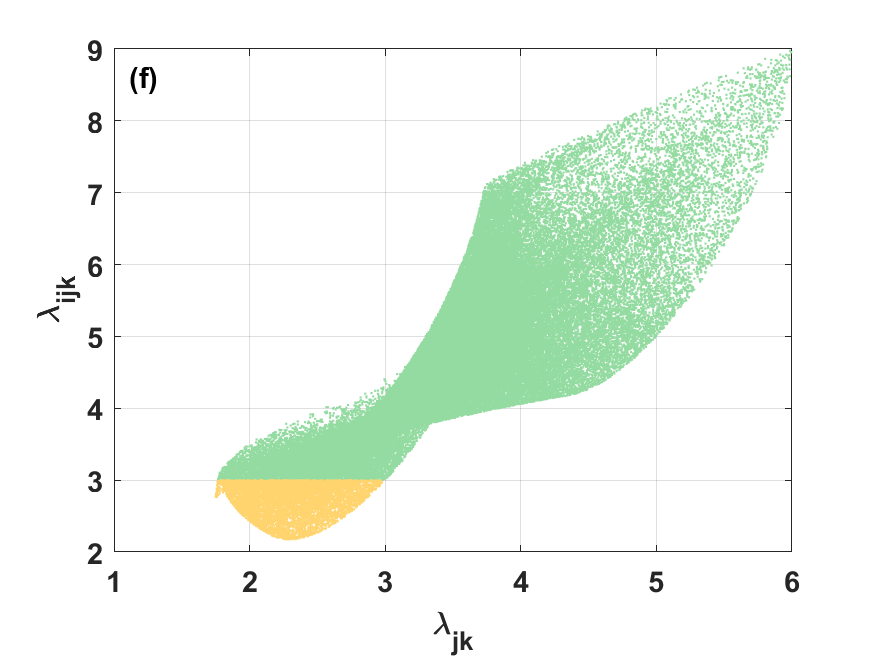}\\
\includegraphics[width=0.42\textwidth]{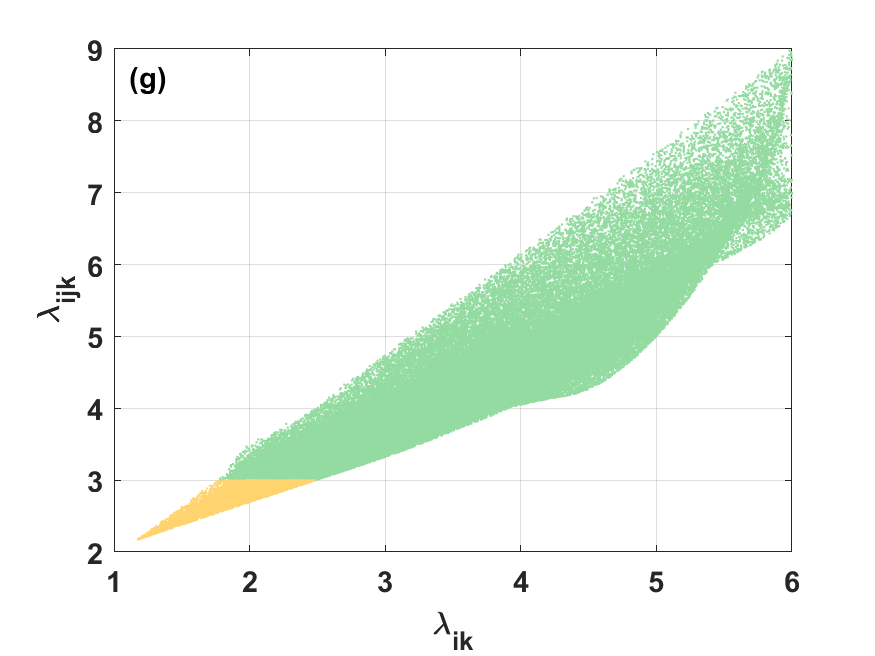}
\caption{\label{Fig_III-2} Graphs revealing the relations among two- and three-mode principal squeeze variances $\lambda_{ij}$, $\lambda_{jk} $, $ \lambda_{ik} $, and $\lambda_{ijk} $ and two- and three-mode negativities $ N_{ij}$, $ N_{jk}$, and $N_{ijk} $ for the states of the type III-2.}
\end{figure}

Let us consider the state classified as subtype III-2:
\begin{equation}
\vert\psi\rangle=C_{000}\vert 000 \rangle +C_{001}\vert 001 \rangle +C_{011}\vert011 \rangle +C_{111}\vert 111\rangle.
\label{eq:psi_000_001_011_111}
\end{equation}
For this state, the following formulas for the two- and three-mode principal squeeze variances are derived:
\begin{eqnarray}
\lambda_{ij}&=&2\left( 1+P_{011}-P_{001}P_{011}+2P_{111}-P_{011}P_{111}-2\sqrt{P_{001}P_{111}}P_{011}\right. \nonumber \\
&&\left.-\mid 2\left(\sqrt{P_{001}P_{111}}\left(1-P_{011}\right)\right)-P_{001}P_{011}-P_{011}P_{111}\mid \right), \nonumber \\
\lambda_{ik}&=&2\left( 1+P_{001}-2P_{000}P_{001}+P_{011}+2P_{111}-P_{011}P_{111}-4\sqrt{P_{000}P_{111}P_{001}P_{011}} \right),\nonumber \\
\lambda_{jk}&=&2\left( 1+P_{001}-P_{000}P_{001}+2P_{011}+2P_{111}-P_{001}P_{011}-2\sqrt{P_{000}P_{011}}P_{001} \right.\nonumber \\
&&\left.-\mid 2\left(\sqrt{P_{000}P_{011}}\left(1-P_{001}\right) \right)-P_{000}P_{001} -P_{001}P_{011}\mid \right)\nonumber \\
\lambda_{ijk}&=& 3 + 2\left(P_{001} - P_{000}P_{001} + 2P_{011} - P_{001}P_{011} + 3P_{111} -P_{011}P_{111} \right) \nonumber \\
&& -4 \left(\sqrt{P_{000}P_{011}} P_{001}  + \sqrt{P_{000}P_{001} P_{011}P_{111}} + \sqrt{P_{001}P_{111}} P_{011} \right) \nonumber \\
&& - 2 \mid 2 \left(\sqrt{P_{000}P_{011}} - \sqrt{P_{000}P_{011}} P_{001} + \sqrt{P_{001}P_{111}} -\sqrt{P_{000}P_{001}P_{011}P_{111}} \right.  \nonumber \\
&&\left. - \sqrt{P_{001}P_{111}} P_{011} \right)-P_{000}P_{001} - P_{001}P_{011}-P_{011}P_{111}\mid .
\label{eq:lambda_III_2}
\end{eqnarray}
Though these formulas are more complex than the previous ones, we observe that all variances can be lower than their standard-quantum limit and the analyzed states are squeezed. A more detailed analysis of the states classified as subtype III-2 is contained in Fig.~\ref{Fig_III-2}.

The relation between the two-mode principal squeeze variances $\lambda_{ij} $ and $ \lambda_{ik} $ and two-mode negativities $ N_{ij} $ and $ N_{ik} $ of the modes being entangled is shown in the graph of Fig.~\ref{Fig_III-2}a. This graph resembles those of Figs.~\ref{Fig_III-1A}a and~\ref{Fig_III-1B}a devoted to the states of types III-1A and III-1B, respectively. The black solid line indicates the limiting values of the squeezing parameters, which are similar to those observed for type III-1A and III-1B states. The minimal and maximal values of the parameters $\lambda_{ij}$ and $\lambda_{ik}$, for given values of negativity, are obtained for the probabilities $P_{\vert 0\rangle_{j}\vert 11\rangle_{ik}}=P_{000}=0$ and $P_{\vert 1\rangle_{k}\vert 00\rangle_{ij}}=P_{111}=0$, respectively. The variance $\lambda_{ij}$ takes the values at the black curve provided that the wave function (\ref{eq:psi_III-2}) is reduced to the form $\vert\psi\rangle=C_{\vert 1\rangle_{k}\vert 00\rangle_{ij}}\vert 1\rangle_{k} \otimes\vert 00\rangle_{ij} +C_{111}\vert 111\rangle$. On the other hand, the variance $\lambda_{ik}$ takes the values at the black curve if the wave function (\ref{eq:psi_III-2}) is reduced to $\vert\psi\rangle=C_{000}\vert 000\rangle +C_{\vert 0\rangle_{j}\vert 11\rangle_{ik}}\vert 0\rangle_{j} \otimes\vert 11\rangle_{ik}$. The smallest value of the principal squeeze variances $\lambda_{ij} = \lambda_{ik} =1.17 $ coincides with that of the states of type III-1.

Following the graph in Fig.~\ref{Fig_III-2}b, the principal squeeze variance $\lambda_{ij}$ is minimal provided that $P_{111}=\frac{1}{4} \left(2-\sqrt{2}\right)\approx 0.15$. On the other hand, the principal squeeze variance $\lambda_{ik}$ attains its minimum when $P_{000}=\frac{1}{4} \left(2+\sqrt{2}\right)\approx 0.85$. If $\lambda_{ij}$ is minimal, two-mode squeezing in other two pairs of modes is not generated ($\lambda_{ik}=\lambda_{jk}\approx 4.29$ similarly as three-mode squeezing ($\lambda_{ijk}=-2 \sqrt{\left(2-\sqrt{2}\right) \left(\frac{1}{4} \left(\sqrt{2}-2\right)+1\right)}-\sqrt{2}+7\approx 4.17$). This behavior resembles that of the type III-1A states. On the other hand, if $\lambda_{ik}$ is minimal, two-mode squeezing in the pair of modes $i-j $ and $j-k $ is not observed ($\lambda_{ij}=\lambda_{jk}=2 \left(\frac{1}{4} \left(-2-\sqrt{2}\right)+2\right)\approx 2.29$), but we have three-mode squeezing ($\lambda_{ijk}=-2 \sqrt{\left(2+\sqrt{2}\right) \left(\frac{1}{4} \left(-2-\sqrt{2}\right)+1\right)}-\sqrt{2}+5 \approx 2.17$). Such behavior resembles that of type III-1B states.

In contrast to the previous cases and following the graph in Fig.~\ref{Fig_III-2}a, there occur additional squeezed states with the negativities $ N_{ij} $ and $ N_{ik} $ not exceeding $0.09$. These states exhibit the strongest two-mode squeezing ($\lambda_{ij} = \lambda_{ik} = 1.75 $) found in non-entangled states.

Squeezing is also found in the pair of modes $j-k $ with zero two-mode negativity $ N_{jk} $, as shown in Figs.~\ref{Fig_III-2}b-c. The yellow area of squeezed states in Fig.~\ref{Fig_III-2}b is analogous to that in Fig.~\ref{Fig_III-1A}b. This reflects the similarity of the relations between the principal squeeze variances $\lambda_{ij}$ and $\lambda_{jk}$ valid for the states of types III-2 and III-1A. The minimum of $\lambda_{jk} = 1.75$ is achieved, as before, for $\lambda_{ij}=2$.

The relation between two-mode squeezing in the pair of modes $i-k $ and the pair of modes $ j-k $ is revealed in the graph of Fig.~\ref{Fig_III-2}c that closely resembles that of Fig.~\ref{Fig_III-1B}b devoted to type III-1B states. Nevertheless, in the graph in Fig.~\ref{Fig_III-2}c we depict also the states exhibiting two-mode squeezing in the pair of modes $j-k$ ($ \lambda_{jk} < 2 $).

The type III-2 states exhibit three-mode squeezing quantified by the three-mode principal squeeze variance $\lambda_{ijk}<3$. The comparison of graphs in Figs.~\ref{Fig_III-1A}c, \ref{Fig_III-1B}c, and \ref{Fig_III-2}d reveals that the range of three-mode negativities $N_{ijk}$ allowing three-mode squeezing is the broadest for the type III-2 states ($ N_{ijk} <0.8$).

Figures~\ref{Fig_III-2}e-~\ref{Fig_III-2}g provide the relation between two- and three-mode squeezing. The yellow area belongs to the states exhibiting three-mode squeezing. Similarly to the states classified as III-1B, the strongest three-mode squeezing for the states of subtype III-2 occurs when the two-mode squeezing is also the strongest for one pair of qubits (namely, qubits $i-k$). On the other hand, the remaining two pairs of qubits do not show squeezing. Importantly, the three-mode principal squeeze variance $\lambda_{ijk}$ reaches values lower than $3$ in various cases in which the two-mode squeezing is observed in one, two, or three pairs of qubits. Surprisingly, three-mode squeezing can also arise in the absence of two-mode squeezing in all three pairs of modes. This resembles the situation found in the states having three-mode entanglement without two-mode entanglement and belonging to subtype III-0. Furthermore, if three-mode squeezing occurs while one of the qubit pairs does not exhibit squeezing, the two-mode principal squeeze variance of this pair remains below $3$.

\subsection{\label{sec:III_3}Type III-3}
Finally, we analyze the states belonging to type III-3, which exhibit both two- and three-mode entanglement among all three modes, i.e $N_{ij}> 0$, $N_{ik}>0$, $N_{jk} > 0$, and $N_{ijk}> 0$ (see Fig.~\ref{Fig1}d). They are parameterized as
\begin{equation}
\vert\psi\rangle=C_{000}\vert 000\rangle +C_{\vert 0\rangle_{k}\vert 11\rangle_{ij}}\vert 0\rangle_{k} \otimes \vert 11\rangle_{ij} +C_{\vert 0\rangle_{j}\vert 11\rangle_{ik}}\vert 0\rangle_{j} \otimes \vert 11\rangle_{ik}
\label{eq:psi_III-3}
\end{equation}
and include, e.g., the states $C_{000}\vert 000 \rangle +C_{101}\vert 101 \rangle +C_{110}\vert 110 \rangle$ and $C_{000}\vert 000 \rangle +C_{011}\vert 011 \rangle+C_{110}\vert 110 \rangle$.

\begin{figure}[ht]
\centering
\includegraphics[width=0.42\textwidth]{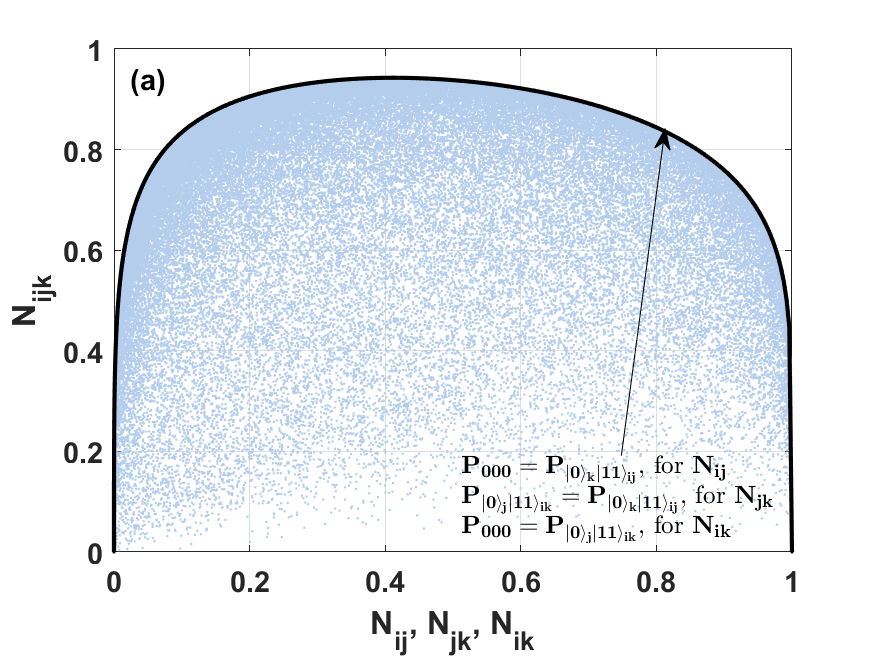}
\includegraphics[width=0.42\textwidth]{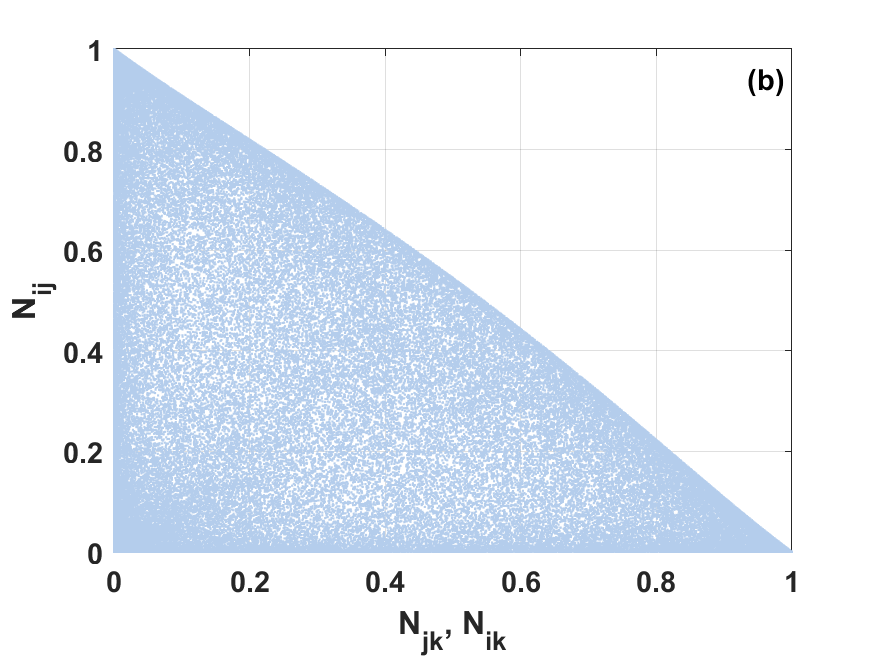}
\caption{\label{Fig_III-3neg} Graphs revealing the relation among three- and two-mode negativities $ N_{ijk}$, $N_{ij} $, $N_{jk} $, and $N_{ik} $ for the states of the type III-3.}
\end{figure}

The relation between two- and three-mode negativities for III-3 states is elucidated in Fig.~\ref{Fig_III-3neg} where the maximal three-mode entanglement is obtained provided that all three two-mode negativities equal to $\frac{1}{3} \left(\sqrt{5}-1\right)\approx 0.41$. However, the three-mode entanglement disappears when one of the two-mode negativities equals 1 and the remaining two-mode negativities are zero. Moreover, if two qubits are maximally entangled, the other pairs of qubits cannot be entangled. In Fig.~\ref{Fig_III-3neg}a, the black curve giving the maximal value of $N_{ijk}$ for the fixed two-mode negativities corresponds to the situation in which two appropriate probabilities equal ($ P_{000}=P_{\vert0\rangle_k\vert11\rangle_{ij}}$). Otherwise, the three-mode negativity  $N_{ijk}$ attains its maxima for fixed $N_{jk}$ ($N_{ik}$) provided that $P_{\vert 0\rangle_j\vert 11\rangle_{ik}}=P_{|0\rangle_k\vert 11\rangle_{ij}}$ ($P_{000}=P_{\vert 0\rangle_j\vert 11\rangle_{ik}}$).

\begin{figure}[ht]
\centering
\includegraphics[width=0.42\textwidth]{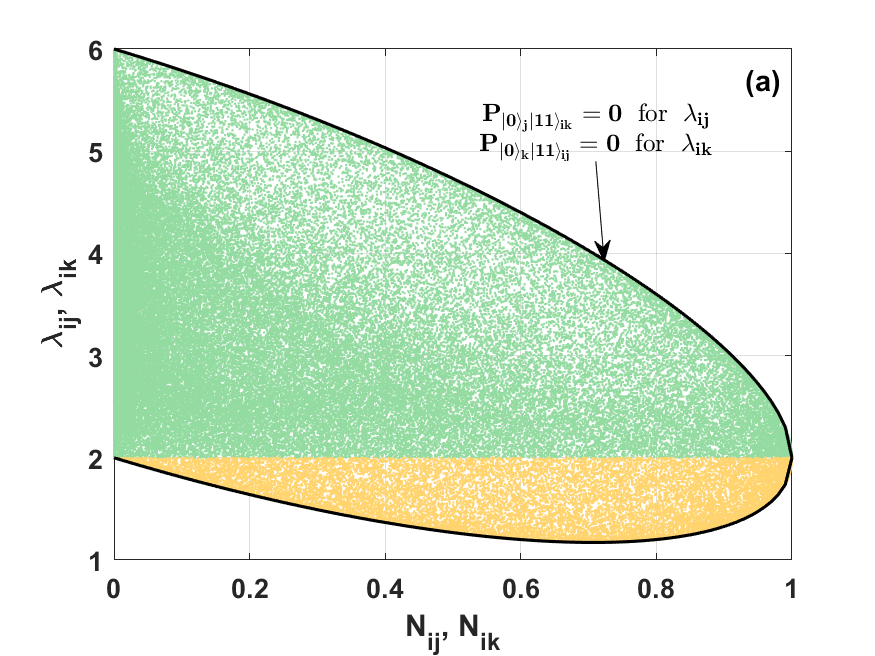}
\includegraphics[width=0.42\textwidth]{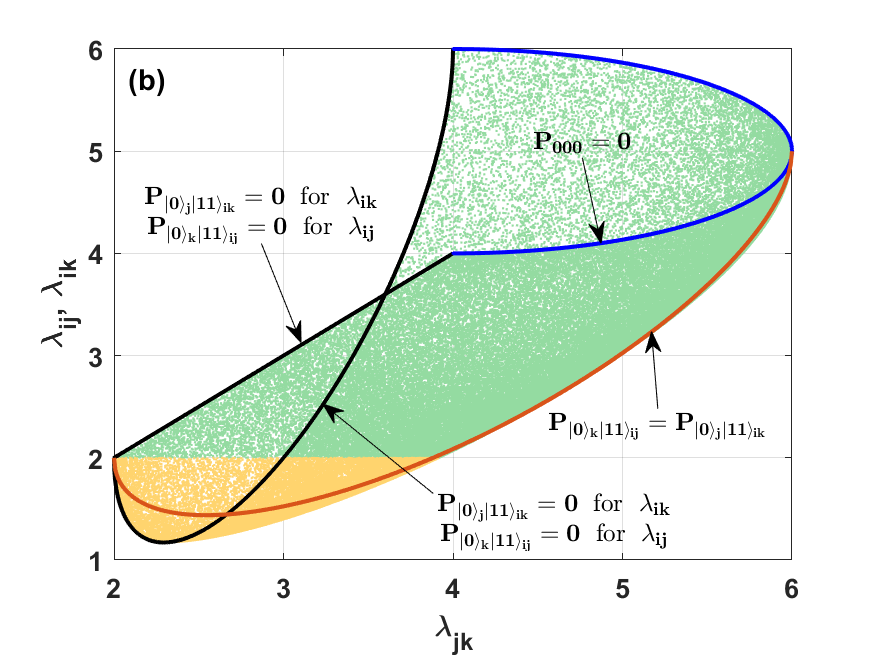}\\
\includegraphics[width=0.42\textwidth]{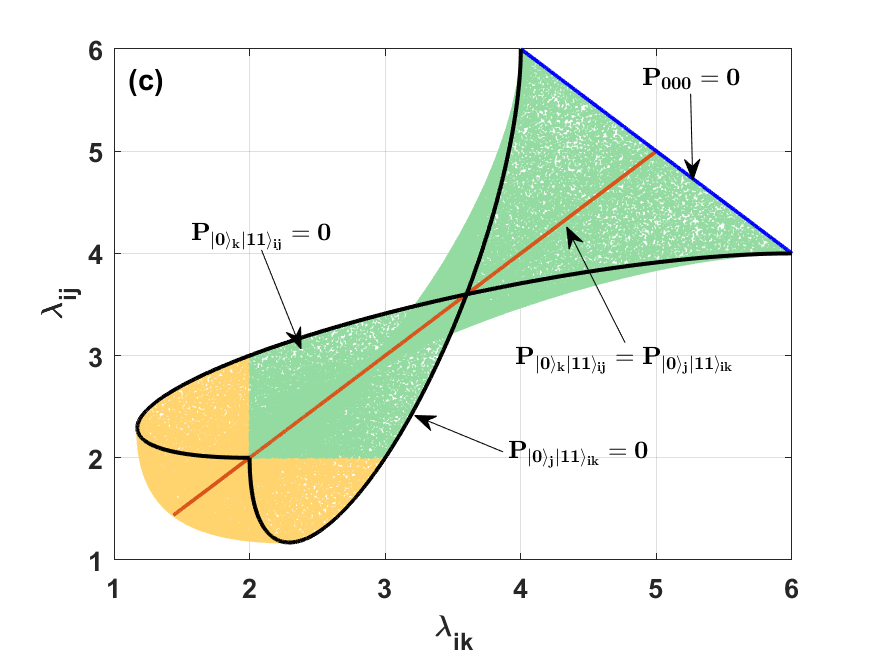}
\includegraphics[width=0.42\textwidth]{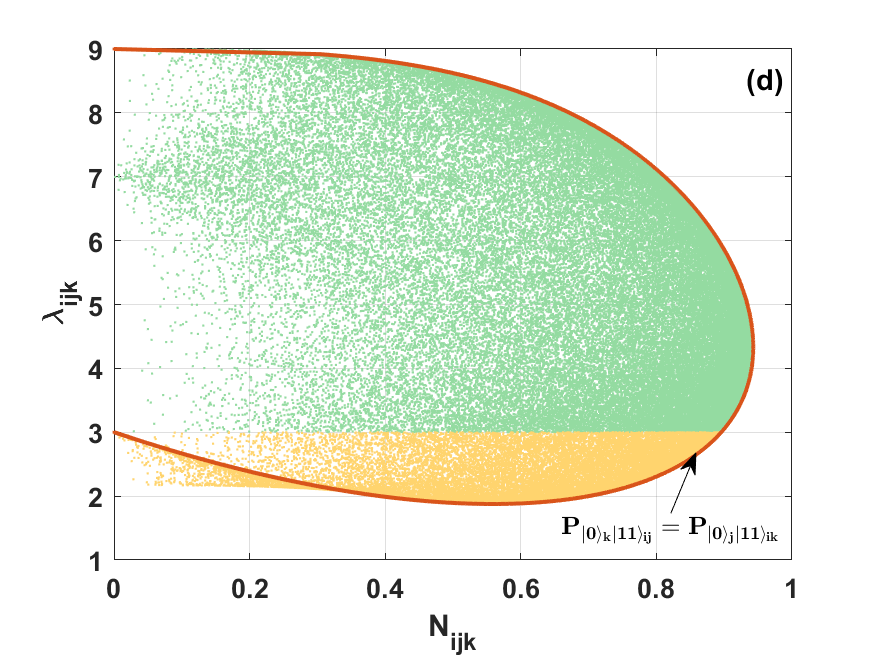}\\
\includegraphics[width=0.42\textwidth]{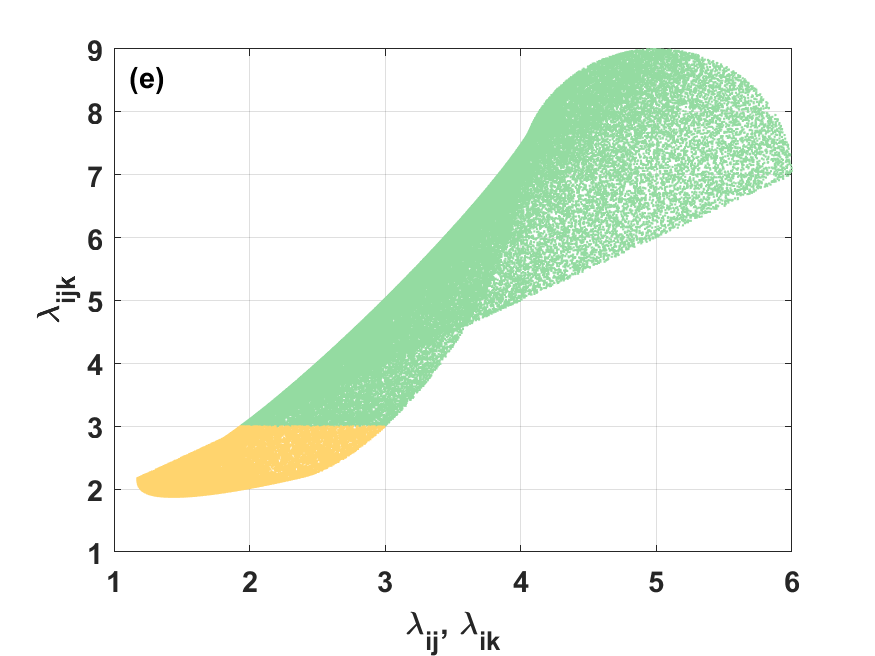}
\includegraphics[width=0.42\textwidth]{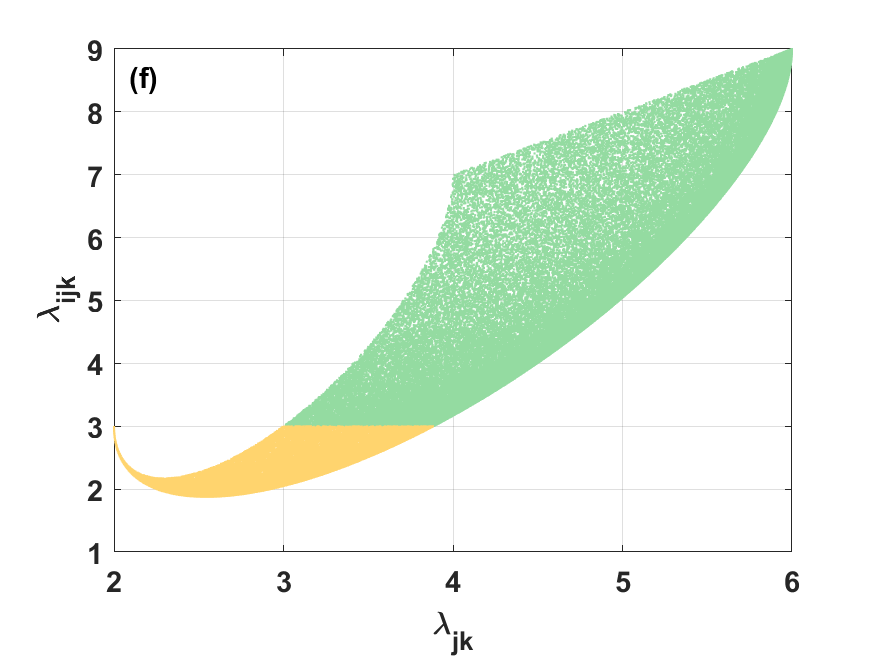}
\caption{\label{Fig_III-3} Graphs revealing the relations among two- and three-mode principal squeeze variances $\lambda_{ij}$, $\lambda_{jk} $, $ \lambda_{ik} $, and $\lambda_{ijk} $ and two- and three-mode negativities $ N_{ij}$, $ N_{ik}$, and $N_{ijk} $ for the states of the type III-3.}
\end{figure}

The states belonging to subtype III-3 include, among others, those described by the following wave function:
\begin{equation}
\vert\psi\rangle=C_{000}\vert 000 \rangle +C_{101}\vert101 \rangle +C_{110}\vert 110\rangle.
\label{eq:psi_000_011_101}
\end{equation}
For these states the principal squeeze variances attain their forms:
\begin{eqnarray}
\lambda_{ij}&=& 2 \left(1 + P_{101} + 2 P_{110} - 2 \sqrt{P_{000}P_{110}} \right), \nonumber \\
\lambda_{ik}&=& 2 \left(1 + P_{110} + 2 P_{101} - 2 \sqrt{P_{000}P_{101}} \right), \nonumber \\
\lambda_{jk}&=& 2 \left(1 + P_{101} + P_{110} + 2 \sqrt{P_{101} P_{110}} \right), \nonumber \\
\lambda_{ijk}&=& 3 + 2 \left(2 P_{101} + 2 P_{110}\right) - 4 \left( \sqrt{P_{000}P_{101}} + \sqrt{P_{000}P_{110}} \right) + 4 \sqrt{P_{101}P_{110}}.
\label{eq:lambda_III_3}
\end{eqnarray}
These formulas express that the three-mode principal squeeze variance can be lower than $3$ on one side, not all two-mode principal squeeze variances can be lower than $2$ on the other side. Further analysis of the states classified as subtype III-3 is provided in Fig.~\ref{Fig_III-3}.

The relation between the two-mode principal squeeze variances $\lambda_{ij} $ and $ \lambda_{ik} $ and the corresponding two-mode negativities $ N_{ij} $ and $ N_{ik} $ plotted in the graph in Fig.~\ref{Fig_III-3}a reflects the corresponding relations found for the III-1A (Fig.~\ref{Fig_III-1A}a) and III-1B (Fig.~\ref{Fig_III-1B}a) states, and also for the III-2 states (Fig.~\ref{Fig_III-2}a).

In Fig.~\ref{Fig_III-3}a similarly as above, the black solid curve determines the minima (maxima) of the principal squeeze variances $\lambda_{ij}$ and $\lambda_{ik}$ reached under the condition $P_{\vert 0\rangle_{j}\vert 11\rangle_{ik}} =0 $ ($P_{\vert 0\rangle_{k}\vert 11\rangle_{ij}}=0$).

The lowest two-mode principal squeeze variances $ \lambda_{ij}=\lambda_{ik} \approx 1.17 $ occur for the two-mode negativities $ N_{ij} = N_{ik} \approx 0.71$ and $P_{000}=\frac{1}{4}\left(2+\sqrt{2}\right)\approx 0.85$. For the lowest value of $\lambda_{ij} $ ($ \lambda_{ik} $), the remaining principal squeeze variances obey $ \lambda_{ik} > 2 $ and $ \lambda_{jk} >2 $ ($\lambda_{ij} > 2 $ and $ \lambda_{jk} >2 $). Interestingly, no squeezing is observed for the pair of modes $j-k$, as illustrated in Fig.~\ref{Fig_III-3}b.

Similarly as type III-1 and type III-2 states, also the type III-3 states exhibit three-mode squeezing, as illustrated in Fig.~\ref{Fig_III-3}d. The range of negativity $ N_{ijk}$ allowing three-mode squeezing is even broader and the three-mode principal squeeze variance $\lambda_{ijk}$ can take on smaller values. The minimal principal squeeze variance $\lambda_{ijk} \approx 1.8$ is found for $N_{ijk}\approx 0.6$. Three-mode squeezed states are observed up to $ N_{ijk} \approx 0.89$.

The relations between the two- and three-mode principal squeeze variances are revealed in Figs.~\ref{Fig_III-3}e and ~\ref{Fig_III-3}f where the yellow area represents the states exhibiting three-mode squeezing ($\lambda_{ijk}<3$). For the III-3 states the largest three-mode squeezing occurs for other states than those endowed with the largest two-mode squeezing. The largest three-mode squeezing is achieved if $\lambda_{ij}=\lambda_{ik}$ attain their smallest possible values. Importantly, the three-mode principal squeeze variance $\lambda_{ijk}$ is smaller than $3$ only when both pairs of qubits exhibit two-mode squeezing, or when only one of them does. Also, when three-mode squeezing is generated, the two-mode principal squeeze variance of the unsqueezed qubit pair is lower than $3$, as noted earlier.

We summarize the above results in Table~\ref{tab1} that qualitatively shows the presence or absence of two- and three-mode entanglement and squeezing in all subgroups of the class III states.
\begin{table}[h]
\caption{Two- and three-mode negativities and principal squeeze variances for the states classified as type III}\label{tab1}%
\begin{tabular}{@{}cccccc@{}}
\toprule
 & III-0  & III-1A & III-1B & III-2 & III-3\\
\midrule
$N_{ij}\neq 0$ &  no  &  no & no & yes & yes \\
$N_{jk}\neq 0$ &  no  &  yes & yes & no & yes \\
$N_{ik}\neq 0$ &  no  &  no & no & yes & yes \\
$N_{ijk}\neq 0$ & yes  &  yes & yes & yes & yes \\
\midrule
$\lambda_{ij}< 2$ &  no  &  yes & no & yes & yes \\
$\lambda_{jk}< 2$ &  no  &  yes & yes &yes & no \\
$\lambda_{ik}< 2$ &  no  &  yes & no & yes & yes \\
$\lambda_{ijk}< 3$ & no  &  yes & yes & yes & yes \\
\botrule
\end{tabular}
\end{table}

\section{\label{sec:conclusions}Conclusions}

In this paper, we have analyzed various three-qubit states that exhibit full three-mode entanglement.

Following the classification of three-qubit states proposed by Sab{\'i}n and Garc{\'i}a-Alcaine in~\cite{SG08}, we have investigated the relations between two- and three-mode entanglement and two- and three-mode squeezing for all four groups of the states exhibiting three-mode entanglement.

We have demonstrated that, except of the GHZ states forming type III-0, the states exhibit both two- and three-mode squeezing. The GHZ states are specific as, though endowed with three-mode entanglement, they do not show any form of two-mode entanglement. For the remaining types of states, we have identified the minima of the two- and three-mode principal squeeze variances in the corresponding space of the states. The same lowest two-mode principal squeeze variances are reached in type III-1, III-2, and III-3 groups of states. On the other hand, the strongest three-mode squeezing is observed for the type III-3 states, that belong to the W-class states. This group of states also allows to observe the three-mode squeezed states that exhibit high level of three-mode entanglement.
Our results show that some states are simultaneously strongly entangled and highly squeezed, both in two- and three-modes. This may seem surprising. However, Wang and Sanders demonstrated in \cite{WS03} that spin squeezing is associated with pairwise entanglement and the increase in squeezing results in the increase of entanglement. Nevertheless, it should be noted that the strongest two-mode entangled states of the analyzed types are characterized by either no squeezing or weak squeezing. In the case of three-mode entanglement, the situation is complex owing to various multipartite entanglement structures and this requires further analysis.

We have revealed that three-mode squeezing occurs alongside with two-mode squeezing found in one, two, or three pairs of qubits. However, for specific states two-mode squeezing is absent. The smallest three-mode principal squeeze variance for the states of types III-1 and III-2 is reached together with the smallest two-mode principal squeeze variances of one or two pairs of qubits.

\noindent\textbf{Acknowledgements} J.K.K. acknowledges the support provided by the program of the National Science Center (NCN) entitled MINIATURA8 program, project no. 2024/08/X/ST2/00753. J.P. acknowledges support by the project No. 25-15775S of the Czech Science Foundation and ITI CZ.02.01.01/00/23\_021/0008790  of the Ministry of Education, Youth, and Sports of the Czech Republic and
EU.

\bibliography{main.bib}

\begin{thebibliography}{10}
\expandafter\ifx\csname url\endcsname\relax
  \def\url#1{\burl{#1}}\fi
\expandafter\ifx\csname urlprefix\endcsname\relax\def\urlprefix{URL }\fi
\providecommand{\bibinfo}[2]{#2}
\providecommand{\eprint}[2][]{\url{#2}}
\providecommand{\doi}[1]{\url{https://doi.org/#1}}
\bibcommenthead

\bibitem{HBV99}
\bibinfo{author}{Hillery, M.}, \bibinfo{author}{Bu\v{z}ek, V.} \&
  \bibinfo{author}{Berthiaume, A.}
\newblock \bibinfo{title}{Quantum secret sharing}.
\newblock \emph{\bibinfo{journal}{Phys. Rev. A}} \textbf{\bibinfo{volume}{59}},
  \bibinfo{pages}{1829--1834} (\bibinfo{year}{1999}).

\bibitem{LSB04}
\bibinfo{author}{Lance, A.~M.}, \bibinfo{author}{Symul, T.},
  \bibinfo{author}{Bowen, W.~P.}, \bibinfo{author}{Sanders, B.~C.} \&
  \bibinfo{author}{Lam, P.~K.}
\newblock \bibinfo{title}{Tripartite quantum state sharing}.
\newblock \emph{\bibinfo{journal}{Phys. Rev. Lett.}}
  \textbf{\bibinfo{volume}{92}}, \bibinfo{pages}{177903}
  (\bibinfo{year}{2004}).

\bibitem{CZZ05}
\bibinfo{author}{Chen, Y.-A.} \emph{et~al.}
\newblock \bibinfo{title}{{Experimental Quantum Secret Sharing and Third-Man
  Quantum Cryptography}}.
\newblock \emph{\bibinfo{journal}{Phys. Rev. Lett.}}
  \textbf{\bibinfo{volume}{95}}, \bibinfo{pages}{200502}
  (\bibinfo{year}{2005}).

\bibitem{K08}
\bibinfo{author}{Kimble, H.~J.}
\newblock \bibinfo{title}{The quantum internet}.
\newblock \emph{\bibinfo{journal}{Nature}} \textbf{\bibinfo{volume}{453}},
  \bibinfo{pages}{1023--1030} (\bibinfo{year}{2008}).

\bibitem{NC10}
\bibinfo{author}{Nielsen, M.~A.} \& \bibinfo{author}{Chuang, I.~L.}
\newblock \emph{\bibinfo{title}{Quantum Computation and Quantum Information:
  10th Anniversary Edition}}  (\bibinfo{publisher}{Cambridge University Press},
  \bibinfo{year}{2010}).

\bibitem{CH16}
\bibinfo{author}{Chitambar, E.} \& \bibinfo{author}{Hsieh, M.-H.}
\newblock \bibinfo{title}{Relating the resource theories of entanglement and
  quantum coherence}.
\newblock \emph{\bibinfo{journal}{Phys. Rev. Lett.}}
  \textbf{\bibinfo{volume}{117}}, \bibinfo{pages}{020402}
  (\bibinfo{year}{2016}).

\bibitem{FML17}
\bibinfo{author}{Figgatt, C.} \emph{et~al.}
\newblock \bibinfo{title}{Complete 3-qubit grover search on a programmable
  quantum computer}.
\newblock \emph{\bibinfo{journal}{Nature Communications}}
  \textbf{\bibinfo{volume}{8}}, \bibinfo{pages}{1918} (\bibinfo{year}{2017}).

\bibitem{CG19}
\bibinfo{author}{Chitambar, E.} \& \bibinfo{author}{Gour, G.}
\newblock \bibinfo{title}{Quantum resource theories}.
\newblock \emph{\bibinfo{journal}{Rev. Mod. Phys.}}
  \textbf{\bibinfo{volume}{91}}, \bibinfo{pages}{025001}
  (\bibinfo{year}{2019}).

\bibitem{WCB23}
\bibinfo{author}{Wang, Y.} \emph{et~al.}
\newblock \bibinfo{title}{An atomic-scale multi-qubit platform}.
\newblock \emph{\bibinfo{journal}{Science}} \textbf{\bibinfo{volume}{382}},
  \bibinfo{pages}{87--92} (\bibinfo{year}{2023}).

\bibitem{GHZ89}
\bibinfo{author}{Greenberger, D.~M.}, \bibinfo{author}{Horne, M.~A.} \&
  \bibinfo{author}{Zeilinger, A.}
\newblock \emph{\bibinfo{title}{Going Beyond Bell's Theorem}},
  \bibinfo{pages}{69--72} (\bibinfo{publisher}{Springer Netherlands},
  \bibinfo{address}{Dordrecht}, \bibinfo{year}{1989}).

\bibitem{DVC00}
\bibinfo{author}{D\"ur, W.}, \bibinfo{author}{Vidal, G.} \&
  \bibinfo{author}{Cirac, J.~I.}
\newblock \bibinfo{title}{Three qubits can be entangled in two inequivalent
  ways}.
\newblock \emph{\bibinfo{journal}{Phys. Rev. A}} \textbf{\bibinfo{volume}{62}},
  \bibinfo{pages}{062314} (\bibinfo{year}{2000}).

\bibitem{ABL01}
\bibinfo{author}{Ac\'{\i}n, A.}, \bibinfo{author}{Bru\ss{}, D.},
  \bibinfo{author}{Lewenstein, M.} \& \bibinfo{author}{Sanpera, A.}
\newblock \bibinfo{title}{Classification of mixed three-qubit states}.
\newblock \emph{\bibinfo{journal}{Phys. Rev. Lett.}}
  \textbf{\bibinfo{volume}{87}}, \bibinfo{pages}{040401}
  (\bibinfo{year}{2001}).

\bibitem{SS07}
\bibinfo{author}{Sharma, S.~S.} \& \bibinfo{author}{Sharma, N.~K.}
\newblock \bibinfo{title}{Two-way and three-way negativities of three-qubit
  entangled states}.
\newblock \emph{\bibinfo{journal}{Phys. Rev. A}} \textbf{\bibinfo{volume}{76}},
  \bibinfo{pages}{012326} (\bibinfo{year}{2007}).

\bibitem{SG08}
\bibinfo{author}{Sab{\'i}n, C.} \& \bibinfo{author}{Garc{\'i}a-Alcaine, G.}
\newblock \bibinfo{title}{A classification of entanglement in three-qubit
  systems}.
\newblock \emph{\bibinfo{journal}{Eur. Phys. J. D}}
  \textbf{\bibinfo{volume}{48}}, \bibinfo{pages}{435--442}
  (\bibinfo{year}{2008}).

\bibitem{EDZ18}
\bibinfo{author}{Enríquez, M.}, \bibinfo{author}{Delgado, F.} \&
  \bibinfo{author}{Życzkowski, K.}
\newblock \bibinfo{title}{Entanglement of three-qubit random pure states}.
\newblock \emph{\bibinfo{journal}{Entropy}} \textbf{\bibinfo{volume}{20}}
  (\bibinfo{year}{2018}).

\bibitem{JPL05}
\bibinfo{author}{Joo, J.}, \bibinfo{author}{Park, Y.-J.}, \bibinfo{author}{Lee,
  J.}, \bibinfo{author}{Jang, J.} \& \bibinfo{author}{Kim, I.}
\newblock \bibinfo{title}{{Quantum Secure Communication via a W State}}.
\newblock \emph{\bibinfo{journal}{J. Korean Phys. Soc.}}
  \textbf{\bibinfo{volume}{46}}, \bibinfo{pages}{763} (\bibinfo{year}{2005}).

\bibitem{DZL06}
\bibinfo{author}{Fu-Guo, D.}, \bibinfo{author}{Ping, Z.},
  \bibinfo{author}{Xi-Han, L.}, \bibinfo{author}{Chun-Yan, L.} \&
  \bibinfo{author}{Hong-Yu, Z.}
\newblock \bibinfo{title}{{Efficient Multiparty Quantum Secret Sharing with
  Greenberger–Horne–Zeilinger States}}.
\newblock \emph{\bibinfo{journal}{Chinese Physics Letters}}
  \textbf{\bibinfo{volume}{23}}, \bibinfo{pages}{1084} (\bibinfo{year}{2006}).

\bibitem{WZT07}
\bibinfo{author}{Jian, W.}, \bibinfo{author}{Quan, Z.} \&
  \bibinfo{author}{Chao-Jing, T.}
\newblock \bibinfo{title}{{Quantum Secure Communication Scheme with W State}}.
\newblock \emph{\bibinfo{journal}{Communications in Theoretical Physics}}
  \textbf{\bibinfo{volume}{48}}, \bibinfo{pages}{637} (\bibinfo{year}{2007}).

\bibitem{TH10}
\bibinfo{author}{Tsai, C.-W.} \& \bibinfo{author}{Hwang, T.}
\newblock \bibinfo{title}{{New deterministic quantum communication via
  symmetric W state}}.
\newblock \emph{\bibinfo{journal}{Optics Communications}}
  \textbf{\bibinfo{volume}{283}}, \bibinfo{pages}{4397--4400}
  (\bibinfo{year}{2010}).

\bibitem{LCW24}
\bibinfo{author}{Li, G.-D.} \emph{et~al.}
\newblock \bibinfo{title}{{Quantum Secret Sharing Enhanced: Utilizing W States
  for Anonymous and Secure Communication}} (\bibinfo{year}{2024}).
\newblock \urlprefix\url{https://arxiv.org/abs/2402.02413}.
\newblock
  \bibinfo{eprint}{{\href{https://arxiv.org/abs/2402.02413}{{arXiv:2402.02413}}}}.

\bibitem{YG01}
\bibinfo{author}{Yang, C.-P.} \& \bibinfo{author}{Gea-Banacloche, J.}
\newblock \bibinfo{title}{Three-qubit quantum error-correction scheme for
  collective decoherence}.
\newblock \emph{\bibinfo{journal}{Phys. Rev. A}} \textbf{\bibinfo{volume}{63}},
  \bibinfo{pages}{022311} (\bibinfo{year}{2001}).

\bibitem{TWJ08}
\bibinfo{author}{Tornberg, L.}, \bibinfo{author}{Wallquist, M.},
  \bibinfo{author}{Johansson, G.}, \bibinfo{author}{Shumeiko, V.~S.} \&
  \bibinfo{author}{Wendin, G.}
\newblock \bibinfo{title}{Implementation of the three-qubit phase-flip error
  correction code with superconducting qubits}.
\newblock \emph{\bibinfo{journal}{Phys. Rev. B}} \textbf{\bibinfo{volume}{77}},
  \bibinfo{pages}{214528} (\bibinfo{year}{2008}).

\bibitem{OV10}
\bibinfo{author}{Ottaviani, C.} \& \bibinfo{author}{Vitali, D.}
\newblock \bibinfo{title}{Implementation of a three-qubit quantum
  error-correction code in a cavity-{QED} setup}.
\newblock \emph{\bibinfo{journal}{Phys. Rev. A}} \textbf{\bibinfo{volume}{82}},
  \bibinfo{pages}{012319} (\bibinfo{year}{2010}).

\bibitem{STC17}
\bibinfo{author}{Sohn, I.}, \bibinfo{author}{Tarucha, S.} \&
  \bibinfo{author}{Choi, B.-S.}
\newblock \bibinfo{title}{Analysis of physical requirements for simple
  three-qubit and nine-qubit quantum error correction on quantum-dot and
  superconductor qubits}.
\newblock \emph{\bibinfo{journal}{Phys. Rev. A}} \textbf{\bibinfo{volume}{95}},
  \bibinfo{pages}{012306} (\bibinfo{year}{2017}).

\bibitem{RDN12}
\bibinfo{author}{Reed, M.~D.} \emph{et~al.}
\newblock \bibinfo{title}{Realization of three-qubit quantum error correction
  with superconducting circuits}.
\newblock \emph{\bibinfo{journal}{Nature}} \textbf{\bibinfo{volume}{482}},
  \bibinfo{pages}{382--385} (\bibinfo{year}{2012}).

\bibitem{LPP88}
\bibinfo{author}{A.~Luk\v{s}, V.~P.} \& \bibinfo{author}{Pe\v{r}ina, J.}
\newblock \bibinfo{title}{{Principal squeezing of vacuum fluctuations}}.
\newblock \emph{\bibinfo{journal}{Opt. Commun.}} \textbf{\bibinfo{volume}{67}},
  \bibinfo{pages}{149} (\bibinfo{year}{1988}).

\bibitem{Dodonov2002}
\bibinfo{author}{Dodonov, V.~V.}
\newblock \bibinfo{title}{Nonclassical states in quantum optics: A squeezed
  review of the first 75 years}.
\newblock \emph{\bibinfo{journal}{J. Opt. B: Quantum Semiclass. Opt.}}
  \textbf{\bibinfo{volume}{4}}, \bibinfo{pages}{R1---R33}
  (\bibinfo{year}{2002}).

\bibitem{Dodonov2003}
\bibinfo{author}{Dodonov, V.~V.} \& \bibinfo{author}{{Man\'ko}, V.~I.}
\newblock \bibinfo{editor}{Dodonov, V.~V.} \& \bibinfo{editor}{{Man\'ko},
  M.~V.} (eds) \emph{\bibinfo{title}{{Nonclassical} states in quantum physics:
  brief historical review}}.
\newblock (eds \bibinfo{editor}{Dodonov, V.~V.} \& \bibinfo{editor}{{Man\'ko},
  M.~V.}) \emph{\bibinfo{booktitle}{Theory of Nonclassical States of Light}},
  \bibinfo{pages}{1---80} (\bibinfo{publisher}{Taylor \& Francis},
  \bibinfo{year}{2003}).

\bibitem{Lvovsky2009}
\bibinfo{author}{Lvovsky, A.~I.} \& \bibinfo{author}{Raymer, M.~G.}
\newblock \bibinfo{title}{Continuous-variable optical quantum state
  tomography}.
\newblock \emph{\bibinfo{journal}{Rev. Mod. Phys.}}
  \textbf{\bibinfo{volume}{81}}, \bibinfo{pages}{299---332}
  (\bibinfo{year}{2009}).

\bibitem{SHY85}
\bibinfo{author}{Slusher, R.~E.}, \bibinfo{author}{Hollberg, L.~W.},
  \bibinfo{author}{Yurke, B.}, \bibinfo{author}{Mertz, J.~C.} \&
  \bibinfo{author}{Valley, J.~F.}
\newblock \bibinfo{title}{{Observation of Squeezed States Generated by
  Four-Wave Mixing in an Optical Cavity}}.
\newblock \emph{\bibinfo{journal}{Phys. Rev. Lett.}}
  \textbf{\bibinfo{volume}{55}}, \bibinfo{pages}{2409--2412}
  (\bibinfo{year}{1985}).

\bibitem{SLW86}
\bibinfo{author}{Shelby, R.~M.}, \bibinfo{author}{Levenson, M.~D.},
  \bibinfo{author}{Walls, D.~F.}, \bibinfo{author}{Aspect, A.} \&
  \bibinfo{author}{Milburn, G.~J.}
\newblock \bibinfo{title}{Generation of squeezed states of light with a
  fiber-optic ring interferometer}.
\newblock \emph{\bibinfo{journal}{Phys. Rev. A}} \textbf{\bibinfo{volume}{33}},
  \bibinfo{pages}{4008--4025} (\bibinfo{year}{1986}).

\bibitem{WKH86}
\bibinfo{author}{Wu, L.-A.}, \bibinfo{author}{Kimble, H.~J.},
  \bibinfo{author}{Hall, J.~L.} \& \bibinfo{author}{Wu, H.}
\newblock \bibinfo{title}{Generation of squeezed states by parametric down
  conversion}.
\newblock \emph{\bibinfo{journal}{Phys. Rev. Lett.}}
  \textbf{\bibinfo{volume}{57}}, \bibinfo{pages}{2520--2523}
  (\bibinfo{year}{1986}).

\bibitem{DDA10}
\bibinfo{author}{Dell'Anno, F.}, \bibinfo{author}{De~Siena, S.},
  \bibinfo{author}{Adesso, G.} \& \bibinfo{author}{Illuminati, F.}
\newblock \bibinfo{title}{Teleportation of squeezing: Optimization using
  non-gaussian resources}.
\newblock \emph{\bibinfo{journal}{Phys. Rev. A}} \textbf{\bibinfo{volume}{82}},
  \bibinfo{pages}{062329} (\bibinfo{year}{2010}).

\bibitem{GH21}
\bibinfo{author}{Goldberg, A.~Z.} \& \bibinfo{author}{Heshami, K.}
\newblock \bibinfo{title}{How squeezed states both maximize and minimize the
  same notion of quantumness}.
\newblock \emph{\bibinfo{journal}{Phys. Rev. A}}
  \textbf{\bibinfo{volume}{104}}, \bibinfo{pages}{032425}
  (\bibinfo{year}{2021}).

\bibitem{RPF22}
\bibinfo{author}{Renger, M.} \emph{et~al.}
\newblock \bibinfo{title}{Flow of quantum correlations in noisy two-mode
  squeezed microwave states}.
\newblock \emph{\bibinfo{journal}{Phys. Rev. A}}
  \textbf{\bibinfo{volume}{106}}, \bibinfo{pages}{052415}
  (\bibinfo{year}{2022}).

\bibitem{GGS24}
\bibinfo{author}{Garc\'{i}a-Beni, J.}, \bibinfo{author}{Giorgi, G.~L.},
  \bibinfo{author}{Soriano, M.~C.} \& \bibinfo{author}{Zambrini, R.}
\newblock \bibinfo{title}{Squeezing as a resource for time series processing in
  quantum reservoir computing}.
\newblock \emph{\bibinfo{journal}{Opt. Express}} \textbf{\bibinfo{volume}{32}},
  \bibinfo{pages}{6733--6747} (\bibinfo{year}{2024}).

\bibitem{NDH25}
\bibinfo{author}{Nguyen, H.~Q.} \emph{et~al.}
\newblock \bibinfo{title}{Digital reconstruction of squeezed light for quantum
  information processing}.
\newblock \emph{\bibinfo{journal}{npj Quantum Information}}
  \textbf{\bibinfo{volume}{11}}, \bibinfo{pages}{71} (\bibinfo{year}{2025}).

\bibitem{MUL12}
\bibinfo{author}{Madsen, L.~S.}, \bibinfo{author}{Usenko, V.~C.},
  \bibinfo{author}{Lassen, M.}, \bibinfo{author}{Filip, R.} \&
  \bibinfo{author}{Andersen, U.~L.}
\newblock \bibinfo{title}{Continuous variable quantum key distribution with
  modulated entangled states}.
\newblock \emph{\bibinfo{journal}{Nature Communications}}
  \textbf{\bibinfo{volume}{3}}, \bibinfo{pages}{1083} (\bibinfo{year}{2012}).

\bibitem{GHD15}
\bibinfo{author}{Gehring, T.} \emph{et~al.}
\newblock \bibinfo{title}{Implementation of continuous-variable quantum key
  distribution with composable and one-sided-device-independent security
  against coherent attacks}.
\newblock \emph{\bibinfo{journal}{Nature Communications}}
  \textbf{\bibinfo{volume}{6}}, \bibinfo{pages}{8795} (\bibinfo{year}{2015}).

\bibitem{ABT24}
\bibinfo{author}{Alexander, B.~J.}, \bibinfo{author}{Bollinger, J.~J.} \&
  \bibinfo{author}{Tame, M.~S.}
\newblock \bibinfo{title}{Robustness of the projected squeezed state protocol}.
\newblock \emph{\bibinfo{journal}{Phys. Rev. A}}
  \textbf{\bibinfo{volume}{109}}, \bibinfo{pages}{052614}
  (\bibinfo{year}{2024}).

\bibitem{HQ23}
\bibinfo{author}{Hillmann, T.} \& \bibinfo{author}{Quijandr\'{\i}a, F.}
\newblock \bibinfo{title}{Quantum error correction with dissipatively
  stabilized squeezed-cat qubits}.
\newblock \emph{\bibinfo{journal}{Phys. Rev. A}}
  \textbf{\bibinfo{volume}{107}}, \bibinfo{pages}{032423}
  (\bibinfo{year}{2023}).

\bibitem{XZW23}
\bibinfo{author}{Xu, Q.} \emph{et~al.}
\newblock \bibinfo{title}{Autonomous quantum error correction and
  fault-tolerant quantum computation with squeezed cat qubits}.
\newblock \emph{\bibinfo{journal}{npj Quantum Information}}
  \textbf{\bibinfo{volume}{9}}, \bibinfo{pages}{78} (\bibinfo{year}{2023}).

\bibitem{KBG24}
\bibinfo{author}{Korolev, S.~B.}, \bibinfo{author}{Bashmakova, E.~N.} \&
  \bibinfo{author}{Golubeva, T.~Y.}
\newblock \bibinfo{title}{Error correction using squeezed {Fock} states}.
\newblock \emph{\bibinfo{journal}{Quantum Information Processing}}
  \textbf{\bibinfo{volume}{23}}, \bibinfo{pages}{354} (\bibinfo{year}{2024}).

\bibitem{P96}
\bibinfo{author}{Peres, A.}
\newblock \bibinfo{title}{Separability criterion for density matrices}.
\newblock \emph{\bibinfo{journal}{Phys. Rev. Lett.}}
  \textbf{\bibinfo{volume}{77}}, \bibinfo{pages}{1413--1415}
  (\bibinfo{year}{1996}).

\bibitem{HHH96}
\bibinfo{author}{Horodecki, M.}, \bibinfo{author}{Horodecki, P.} \&
  \bibinfo{author}{Horodecki, R.}
\newblock \bibinfo{title}{Separability of mixed states: Necessary and
  sufficient conditions}.
\newblock \emph{\bibinfo{journal}{Phys. Lett. A}}
  \textbf{\bibinfo{volume}{223}}, \bibinfo{pages}{1--8} (\bibinfo{year}{1996}).

\bibitem{KCH22}
\bibinfo{author}{Kumar, S.} \emph{et~al.}
\newblock \bibinfo{title}{Tripartite entanglement in quantum memristors}.
\newblock \emph{\bibinfo{journal}{Phys. Rev. Appl.}}
  \textbf{\bibinfo{volume}{18}}, \bibinfo{pages}{034004}
  (\bibinfo{year}{2022}).

\bibitem{KP97}
\bibinfo{author}{Korolkova, N.} \& \bibinfo{author}{Pe\v{r}ina, J.}
\newblock \bibinfo{title}{Quantum statistics and dynamics of {Kerr} nonlinear
  couplers}.
\newblock \emph{\bibinfo{journal}{Opt. Commun.}}
  \textbf{\bibinfo{volume}{136}}, \bibinfo{pages}{135--149}
  (\bibinfo{year}{1996}).

\bibitem{AP00}
\bibinfo{author}{Ariunbold, G.} \& \bibinfo{author}{Pe\v{r}ina, J.}
\newblock \bibinfo{title}{Quantum statistics of contradirectional {Kerr}
  nonlinear couplers}.
\newblock \emph{\bibinfo{journal}{Opt. Commun.}}
  \textbf{\bibinfo{volume}{176}}, \bibinfo{pages}{149--154}
  (\bibinfo{year}{2000}).

\bibitem{LK11}
\bibinfo{author}{Leo\'nski, W.} \& \bibinfo{author}{Kowalewska-Kud{\l}aszyk,
  A.}
\newblock \bibinfo{title}{ in \textit{Quantum scissors -- finite-dimensional
  states engineering}} (ed.\bibinfo{editor}{Wolf, E.})
  \emph{\bibinfo{booktitle}{Progress in Optics}}, Vol.~\bibinfo{volume}{56}
  \bibinfo{pages}{131--185} (\bibinfo{publisher}{Elsevier},
  \bibinfo{year}{2011}).

\bibitem{KCL05}
\bibinfo{author}{Korbicz, J.~K.}, \bibinfo{author}{Cirac, J.~I.} \&
  \bibinfo{author}{Lewenstein, M.}
\newblock \bibinfo{title}{Spin squeezing inequalities and entanglement of $n$
  qubit states}.
\newblock \emph{\bibinfo{journal}{Phys. Rev. Lett.}}
  \textbf{\bibinfo{volume}{95}}, \bibinfo{pages}{120502}
  (\bibinfo{year}{2005}).

\bibitem{MWS11}
\bibinfo{author}{Ma, J.}, \bibinfo{author}{Wang, X.}, \bibinfo{author}{Sun, C.}
  \& \bibinfo{author}{Nori, F.}
\newblock \bibinfo{title}{Quantum spin squeezing}.
\newblock \emph{\bibinfo{journal}{Physics Reports}}
  \textbf{\bibinfo{volume}{509}}, \bibinfo{pages}{89--165}
  (\bibinfo{year}{2011}).

\bibitem{G12}
\bibinfo{author}{Gross, C.}
\newblock \bibinfo{title}{Spin squeezing, entanglement and quantum metrology
  with bose–einstein condensates}.
\newblock \emph{\bibinfo{journal}{Journal of Physics B: Atomic, Molecular and
  Optical Physics}} \textbf{\bibinfo{volume}{45}}, \bibinfo{pages}{103001}
  (\bibinfo{year}{2012}).

\bibitem{KKJ19}
\bibinfo{author}{Kalaga, J.~K.}, \bibinfo{author}{Kowalewska-Kud{\l}aszyk, A.},
  \bibinfo{author}{Jarosik, M.~W.}, \bibinfo{author}{Szcz\c{e}\'{s}niak, R.} \&
  \bibinfo{author}{Leo\'nski, W.}
\newblock \bibinfo{title}{Enhancement of the entanglement generation via
  randomly perturbed series of external pulses in a nonlinear bose–hubbard
  dimer}.
\newblock \emph{\bibinfo{journal}{Nonlinear Dynamics}}
  \textbf{\bibinfo{volume}{97}}, \bibinfo{pages}{1619--1633}
  (\bibinfo{year}{2019}).

\bibitem{KK19}
\bibinfo{author}{Kalaga, J.~K.} \& \bibinfo{author}{Kowalewska-Kud{\l}aszyk,
  A.}
\newblock \bibinfo{title}{Frequency variations in impulse excitations as a way
  of entanglement increase in the two-mode bose-hubbard model}.
\newblock \emph{\bibinfo{journal}{J. Opt. Soc. Am. B}}
  \textbf{\bibinfo{volume}{36}}, \bibinfo{pages}{2140--2146}
  (\bibinfo{year}{2019}).

\bibitem{NF25}
\bibinfo{author}{Nair, J. M.~P.} \& \bibinfo{author}{Flebus, B.}
\newblock \bibinfo{title}{Engineering long-lived entanglement through
  dissipation in quantum hybrid solid-state platforms} (\bibinfo{year}{2025}).
\newblock \urlprefix\url{https://arxiv.org/abs/2410.15588}.
\newblock
  \bibinfo{eprint}{{\href{https://arxiv.org/abs/2410.15588}{{arXiv:2410.15588}}}}.

\bibitem{WS03}
\bibinfo{author}{Wang, X.} \& \bibinfo{author}{Sanders, B.~C.}
\newblock \bibinfo{title}{Spin squeezing and pairwise entanglement for
  symmetric multiqubit states}.
\newblock \emph{\bibinfo{journal}{Phys. Rev. A}} \textbf{\bibinfo{volume}{68}},
  \bibinfo{pages}{012101} (\bibinfo{year}{2003}).

\end{thebibliography}

\end{document}